\documentclass[tighten,twocolumn,appendixfloats,apj]{likeapj}
\pdfoutput=1

\usepackage{etoolbox}

\makeatletter

\patchcmd{\NAT@citex}
  {\@citea\NAT@hyper@{%
     \NAT@nmfmt{\NAT@nm}%
     \hyper@natlinkbreak{\NAT@aysep\NAT@spacechar}{\@citeb\@extra@b@citeb}%
     \NAT@date}}
  {\@citea\NAT@nmfmt{\NAT@nm}%
   \NAT@aysep\NAT@spacechar\NAT@hyper@{\NAT@date}}{}{}

\patchcmd{\NAT@citex}
  {\@citea\NAT@hyper@{%
     \NAT@nmfmt{\NAT@nm}%
     \hyper@natlinkbreak{\NAT@spacechar\NAT@@open\if*#1*\else#1\NAT@spacechar\fi}%
       {\@citeb\@extra@b@citeb}%
     \NAT@date}}
  {\@citea\NAT@nmfmt{\NAT@nm}%
   \NAT@spacechar\NAT@@open\if*#1*\else#1\NAT@spacechar\fi\NAT@hyper@{\NAT@date}}
  {}{}

  \makeatother
\usepackage{graphicx,bm,mathptmx,amsmath,float}

\hypersetup{linkcolor=blue,citecolor=blue,filecolor=cyan,urlcolor=blue,bookmarksnumbered=true,bookmarksopen=true}
\usepackage[all]{hypcap}
\usepackage{natbib}
\usepackage{graphicx}
\usepackage{epsfig}
\usepackage{array}
\usepackage{multirow}
\newcolumntype{C}[1]{>{\centering}m{#1}}

\usepackage{subscript}

\usepackage{textgreek} %no math mode - just \textalpha etc
\usepackage{upgreek} %in math mode - just $\upalpha$ etc

\newcommand{\kmps}{km s$^{-1}$}

\newcommand{\mcgas}{$M_{\rm CGas}$}
\newcommand{\mhgas}{$M_{\rm HGas}$}

\newcommand{\mstar}{$M_{\ast}$}
\newcommand{\msun}{$M_{\odot}$}

\newcommand{\dda}{$\Delta D_{\rm A}$}

\newcommand{\mhone}{$M_{\rm H{\textsc i}}$}
\newcommand{\mhtwom}{$M_{\rm H_2}$}

\newcommand{\x}{X-ray}

\newcommand{\hone}{H{\sc \,i}}

\newcommand{\htwom}{H$_2$}

\newcommand{\galex}{{\it GALEX}}

\newcommand{\hst}{{\it HST}}

\newcommand{\spitzer}{{\it Spitzer}}

\newcommand{\swift}{{\it Swift}}

\newcommand{\wone}{{\it uvw1}}
\newcommand{\wtwo}{{\it uvw2}}

\newcommand{\mtwo}{{\it uvm2}}
\newcommand{\uvot}{{UVOT}}

\newcommand{\clues}{{\sc clues}}

\newcommand{\fr}{Figure~\ref}

\newcommand{\scr}{Section~\ref}
\newcommand{\tr}{Table~\ref}
\newcommand{\exi}{\begin{equation}}
\newcommand{\exo}{\end{equation}}
\newcommand{\osp}{\hspace{4.5pt}} %small positive space equal to $1$
\newcommand{\psp}{\hspace{7pt}} %small positive space equal to $-$
\newcommand{\tm}{\tablenotemark}

\newcommand{\txs}{\textsuperscript}
\newcommand{\mic}{$\upmu$m}

\def\spose#1{\hbox to 0pt{#1\hss}} 
\def\approxlt{\mathrel{\spose{\lower 3pt\hbox{$\sim$}}
        \raise 2.0pt\hbox{$<$}}}
\def\approxgt{\mathrel{\spose{\lower 3pt\hbox{$\sim$}}
        \raise 2.0pt\hbox{$>$}}}

\usepackage{subfigure}
\newcolumntype{M}{>{$}c<{$}}
\newcommand{\rah}{$^{\rm h}$}
\newcommand{\ram}{$^{\rm m}$}
\newcommand{\ras}{$^{\rm s}$}
\newcommand{\dtheta}{$\Delta \theta_{\rm vel}$}
\newcommand{\dd}{$\Delta D$}

\newcommand{\ms}{Millennium Simulation}
\newcommand{\mms}{milli-Millennium Simulation}
\newcommand{\mmsg}{mGuo2010a}

\received{2018 November 6}
\revised{2018 December 28}
\accepted{2018 December 31}
\published{2019 February 5}
\shorttitle{}
\shortauthors{Tzanavaris et al.}

\begin{document}

\title{\hspace{-1cm}Cosmic Pathways for Compact Groups in the Milli-Millennium Simulation}

\author{P.~Tzanavaris}%\altaffiliation{Not-A Fellow}
\affil{University of Maryland, Baltimore County, 1000
  Hilltop Circle, Baltimore, MD 21250, USA}
\affil{Astrophysics Science Division, Laboratory for X-ray Astrophysics, NASA/Goddard Space Flight Center, Mail Code 662, Greenbelt, MD, 20771, USA
}
\author{S.~C.~Gallagher}
\affil{Department of Physics and Astronomy, University of Western Ontario, London, ON N6A 3K7, Canada}
\affil{Centre for Planetary and Space Exploration, University of Western Ontario, London, ON N6A 5B9, Canada}
\affil{Rotman Institute of Philosophy, University of Western Ontario, London, ON N6A 5B9, Canada}
\affil{Canadian Space Agency, Saint-Hubert, QC J3Y 8Y9, Canada}
\author{S.~Ali}
\affil{Department of Electrical and Computer Engineering, University of Western Ontario, London, ON N6A 5B9, Canada}
\affil{InademyComputing: Research, Development and Consultancy, London, ON N6A 1C9, Canada}

\author{D.~R.~Miller}
\affil{Department of Physics and Astronomy, University of Western Ontario, London, ON N6A 3K7, Canada}
\affil{Department of Physics, Simon Fraser University, Burnaby, BC V5A 1S6, Canada}

\author{S.~Pentinga}
\affil{Department of Physics and Astronomy, University of Western Ontario, London, ON N6A 3K7, Canada}

\author{K.~E.~Johnson}
\affil{Department of Astronomy, University of Virginia, P.O. Box 400325, Charlottesville, VA 22904-4325, USA}
\affil{National Radio Astronomy Observatory, 520 Edgemont Road, Charlottesville, VA 22903, USA}

\begin{abstract}
We detected 10 compact galaxy groups (CGs) at $z=0$ in the semianalytic galaxy catalog of \citet{guo2011} for the milli-Millennium Cosmological Simulation (sCGs in \mmsg).  We aimed to identify potential canonical pathways for compact group evolution and thus illuminate the history of observed nearby compact groups. By constructing  merger trees for $z=0$ sCG galaxies, we studied the cosmological evolution of key properties, and compared them with $z=0$ Hickson CGs (HCGs). We found that, once sCG galaxies come within 1 (0.5) Mpc of their most massive galaxy, they remain within that distance until $z=0$, suggesting sCG ``birth redshifts.'' At $z=0$ stellar masses of sCG most massive galaxies are within $10^{10} \lesssim M_{\ast}/M_{\odot} \lesssim 10^{11}$. In several cases, especially in the two four- and five-member systems, the amount of cold gas mass anticorrelates with stellar mass, which in turn correlates with hot gas mass.  We define the angular difference between group members' 3D velocity vectors,  $\Delta\theta_{\rm vel}$, and note that many of the groups are long-lived because their small values of $\Delta\theta_{\rm vel}$ indicate a significant parallel component.  For triplets in particular,  $\Delta\theta_{\rm vel}$ values range between $20\degr$ and $40\degr$ so that galaxies are coming together along roughly parallel paths, and pairwise separations do not show large pronounced changes after close encounters. The best agreement between sCG and HCG physical properties is for \mstar\ galaxy values, but HCG values are higher overall, including for star formation rates (SFRs). Unlike HCGs, due to a tail at low SFR and \mstar, and a lack of \mstar~$\gtrsim 10^{11}$~\msun\ galaxies, only a few sCG galaxies are on the star-forming main sequence.

\end{abstract}

\section{Introduction}
As their name implies, compact groups of galaxies (CGs) constitute a distinct class of galaxy systems, consisting of agglomerations of just a few galaxies, typically separated by a few galaxy radii (median projected separations of $\sim 40 \ h^{-1}$ kpc, where $h\equiv H_0/100$ and $H_0$ is the Hubble constant at $z=0$). Their low galaxy velocity dispersions (radial median of $\sim$200~\kmps), combined with high galaxy number densities \citep[up to $10^8 h^2 \, \rm Mpc^{-2}$;][]{hickson1982,hickson1992}, make them an environment favoring strong and prolonged galaxy interactions. Understanding galaxy evolution in CGs is of great interest, as most galaxies spend the majority of their time in some type of group environment \citep[][and references therein]{mulchaey2000,karachentsev2005}.

Since the compilation of dedicated observational CG catalogs more than 30 years ago \citep{rose1977,hickson1982}, it has become clear that galaxies in the CG environment show characteristics that are distinct compared to virtually all other extragalactic environments, including field galaxies, isolated pairwise mergers, and galaxy clusters \citep{johnson2007,tzanavaris2010,walker2010,walker2012,lenkic2016,zucker2016}, with the notable exception of the Coma Cluster outskirts \citep{walker2010,walker2012}.
CG galaxies show a ``canyon" in {\it mid-infrared} (MIR) color space, suggestive of a rapid transition from actively star-forming to quiescent systems, although most CG galaxies reside in the {\it optical} ``red sequence" rather than the ``green valley" \citep{walker2013,zucker2016}. Consistent with the latter observation, and compared to loose groups and field galaxies, late-type CG galaxies have overall markedly reduced stellar-mass normalized star formation rates (SFRs) (``specific" SFRs [sSFRs]; \citealt{coenda2015}) and a higher fraction of quiescent galaxies that are not actively forming stars \citep{coenda2015,lenkic2016}. \citet{tzanavaris2016} found that, compared to general correlations established for late-type galaxies in the local universe \citep[e.g.,][]{mineo2012}, many of these late-type galaxies exhibit a pronounced excess in \x\ luminosity from high-mass \x\ binaries for their SFR. The MIR canyon is further consistent with an observed bimodality in sSFRs, which is essentially exclusive to CGs \citep{tzanavaris2010,lenkic2016}, with some similar behavior observed in loose galaxy groups \citep{wetzel2012}. Further, by studying warm \htwom, \citet{alatalo2015} also observed bimodal suppression of star formation \citep[see also][]{cluver2013}.

%\begin{minipage}[l]{\linewidth}{$\dagger$ Authorship statement:
%\end{minipage}}\\

CG galaxies tend to have older stellar populations on average \citep{proctor2004,mendesdeoliveira2005,coenda2015}. Compared to loose groups, they correspondingly show a larger fraction of red and early-type systems \citep{coenda2012}. Their brightest group galaxies are more luminous, more massive, larger, redder, and are more frequently classified as ellipticals than the other galaxies in the groups \citep{martinez2013}.

Overall, these features of CGs compared to other environments imply that the CG environment affects member galaxy evolution in a unique way, possibly linked to pronounced interaction activity that is likely to be occurring, or has occurred in the past, in these systems \citep[see the Fabry-P{\'e}rot work in][]{plana1998,mendes1998,mendes2003,amram2004,torres-flores2009,torres-flores2010,torres-flores2014}. 

For the most part, observed CGs \citep[e.g.,][]{hickson1992,barton1996,lee2004,deng2007,mcconnachie2009,diaz-gimenez2012} only allow us to study the effects of the CG environment in the relatively nearby universe. For instance, the Redshift Survey Compact Group catalog \citep{barton1996} reaches $m_B < 15.5$ and $z\lesssim 0.03$. The advent of cosmological simulations in the past few decades has provided a novel way to probe the evolution of structure in the universe by comparing observations with simulations. In the case of CGs, one can compare observations with simulations at $z=0$ to identify simulated systems that may resemble observed ones. With the information from the simulation database in hand, one can then look backward in time along the merger trees of the ``best-fit'' simulated systems and learn about plausible evolutionary histories for individual CGs.

A number of studies have used galaxy catalogs derived from the dark matter Millennium Simulation \citep{springel2005} to study CGs. \citet{mcconnachie2008} used a mock galaxy catalog derived using the code of \citet{blaizot2005} from the \citet{delucia2007} semianalytic galaxy catalogs. They identified mock CGs in projection in the \ms\ and compared them to observed Hickson CGs (HCGs), finding that about a third of the mock CGs are truly physically dense systems of three or more galaxies, while the remaining are chance alignments. \citet{diaz-gimenez2010} further explored the origin of this issue, finding fractions of physically dense systems between 20\%\ and 45\%, depending on the semianalytic galaxy catalog used. On the other hand, \citet{diaz-gimenez2015} compared CGs from the 2MASS catalog to those in the \citet{henriques2012} mock light cone, whose galaxy properties were produced via the \citet{guo2011} semianalytic model. They found that about two-thirds of CGs are truly isolated and not part of larger structures.
%REF
\citet{snaith2011} used four semianalytic models to compare the predicted luminosity distribution of both loose and compact groups among different models and observations. In spite of some agreement with observations, they found clear differences between semianalytic models. \citet{farhang2017} used the \citet{delucia2007} semianalytic galaxy catalog of the Millennium Simulation to compare the mass assembly histories of compact and fossil groups. They found no strong evidence that CGs are part of a general evolutionary group path that leads to the eventual formation of fossil groups. They concluded that CGs should instead constitute a specific class of groups with a distinct evolutionary path.

In this paper, we take an alternative approach and look for small systems of galaxies that are physically close in 3D space, rather than apparently close in a 2D projection that approximates an astronomical image.  In addition, we use a subset of the galaxy catalog of \citet{guo2011}  known as the milli-Millennium simulation.
%\todon{Unless my understanding is wrong, I believe their model was done on the full \ms, but only the \mms\ data is open access - yes, so rephrased.}
We refer to this subset of the catalog as \mmsg\ 
%{\bf SCG: I suggest we refer to this as mGuo2010a rather than Guo2010a, because Guo et al. was not done on the mMS, but on the whole thing.}, 
and use it to search for CGs forming in the simulation by means of a clustering algorithm to identify small systems of galaxies in physically close proximity that are also isolated. We stress that, because the volume of the simulation is very small (only 1/512 of the full \ms\ volume), we are not looking for statistical samples and patterns.  Instead, we are examining in detail the small number of simulation compact groups (sCGs) in \mmsg\ and following the evolution of their individual components through space and time.  We then compare the properties of the simulated galaxies at $z=0$ with well-known CG members observed in the local universe to look for the best matches.  The aim is to search for evolutionary analogs to elucidate the possible histories of observed CGs.  Rather than making any sweeping claims on CG formation and evolution as a whole, we simply treat the sCGs as interesting ``case studies" to help establish and explore the potential parameter space of CG histories and set up future studies.

%Continue footnote numbering after author affiliations per ApJ style:
\setcounter{footnote}{11}
  The structure of the paper is as follows: \scr{sec-ident} describes our strategy and algorithm for identifying CGs in the simulation, visualization, and results, including key properties of identified systems. \scr{sec-compare} compares \mmsg-detected sCGs to observed HCGs. \scr{sec-disc} discusses our findings. The paper concludes with a summary in \scr{sec-summ}. Consistent with the \ms, we make use of cosmological parameters $H_0=73 \ {\rm km} \ {\rm s}^{-1} \ {\rm Mpc}^{-1}$, $\Omega_M=0.25$, and $\Omega_\Lambda=0.75$ throughout. This is also consistent with observational results from the Carnegie Hubble Program \citep{freedman2012}, as well as the WMAP \citep{hinshaw2013} mission\footnote{All distances and masses originating in simulations are scaled to $h=1$.}.

%\vspace{-0.27cm}
\section{Compact Groups in the \citet{guo2011} Catalog of the Milli-Millennium Simulation}\label{sec-ident}
\subsection{Simulation}

%%%%%%%%%%%%%%%%%%%%%%%%%%%%%%%%%%%%%%%%%%%%%%%%%%%%%%%%%%%%%%%%%%%%%%%%%%%%%%%%%%%%%%%%%%%%%%%%%%%%%%%%%%%%%%%%%
The Millennium Cosmological Simulation \citep{springel2005} is one of the best-known high-performance computing simulations tracing the evolution of dark matter structure over cosmic time.
It was one of the first computer simulations that provided us with a detailed look at the universe at large scales while still being able to resolve structures smaller than the size of individual galaxies. % \citep{springel2005}.
The simulation used $2160^3$ particles of mass $8.6\times 10^8 \ h^{-1} \ M_{\odot}$ inside a comoving cube $500 \ {\rm Mpc} \ h^{-1}$ on a side. It traces the evolution of this cube in 64 redshift snapshots between $z=127$ and $z=0$, a range of approximately 13.56 Gyr in cosmic time. %\citep{wright2006}.
\fr{fig-millennium_sim} displays a snapshot of a section of the \ms\ at $z=0$.

\renewcommand{\baselinestretch}{0.6}\selectfont
\begin{figure}
 \centering
    \includegraphics[scale=.24]{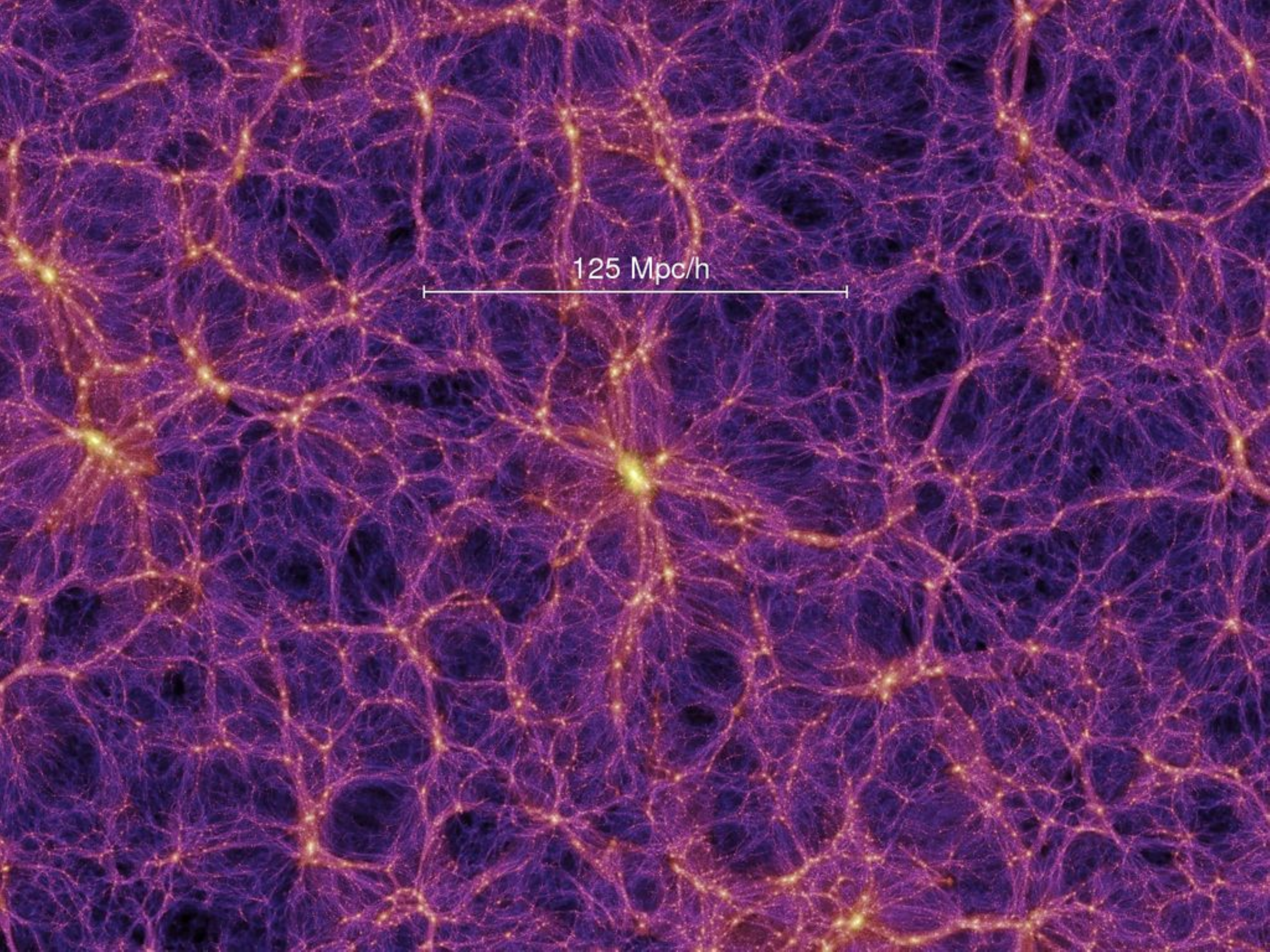}
\caption{Snapshot of the projected dark matter density field for a $15 \ {\rm Mpc} \ h^{-1}$ thick slice of the \ms\ at $z=0$ \citep{springel2005}. The $125 \ {\rm Mpc} \ h^{-1}$ scale bar indicates twice the length of the \mms\ box.}
\label{fig-millennium_sim}
\end{figure}

\renewcommand{\baselinestretch}{1}\selectfont

To compare with observed galaxies, dark matter halos are first identified in the simulations, and subsequently baryonic galaxies with their associated properties are assigned to them by means of semianalytic techniques \citep[e.g.,][for the \ms]{delucia2007,guo2010,guo2011,guo2013}.
\citet{guo2011} have applied their semianalytic method to produce galaxy catalogs associated with the full \ms.  A chunk of the volume in the full simulation is also available, namely, the \mms. This second catalog, \mmsg\footnote{\href{http://gavo.mpa-garching.mpg.de/Millennium/Help/databases/millimil/guo2010a}{http:~//gavo.mpa-garching.mpg.de~/~Millennium~/~Help~/~databases~/~millimil~/\linebreak guo2010a}. Note that the reference for database mGuo2010a is \citet{guo2011}. }, is available for testing of analysis that can then be applied to the full simulation after successful verification.  The smaller volume of this catalog is more appropriate for the particular clustering detection algorithm used in this paper.  We are not searching for statistical samples that would require the full \ms\ volume, but rather interesting individual case studies that allow us to investigate diverse samples of CG evolution over time.  

The \mms\ contains information for the same redshift snapshots as the full simulation but in a cube $62.5 \ {\rm Mpc} \ h^{-1}$ on a side, resulting in a volume approximately $1/512$ that of the full simulation. Many well-studied HCGs are located within 60~Mpc of the Milky Way.  Therefore, within a volume of this size, one would expect to find examples in the simulation that are similar to observed CGs if the simulation reasonably represents our universe and the volume around the Local Group is not unusual.

\begin{deluxetable*}{cccccMccccM}[ht]
  \vspace{-0.5cm}
%\tablewidth{0}
%use the following array stretch command to scale the graph vertically to fit
%use the later \hspace commands to stretch the horizontal white space for appropriate fit
\renewcommand{\arraystretch}{.70}
\tablecaption{Milli-Millennium Simulation Compact Groups \label{tab-simdata}}
\tablehead{
\multicolumn{2}{c}{\hspace{.5cm}ID}\hspace{.5cm}
& \colhead{\hspace{.9cm}$M_{*,\ z=0}$}\hspace{.4cm}
& \colhead{\hspace{.4cm}$M_{{\rm C Gas}, \ z=0}$}\hspace{.4cm}
& \colhead{\hspace{.4cm}$M_{{\rm H Gas}, \ z=0}$}\hspace{.4cm}
& \colhead{\hspace{.5cm}SFR$_{z=0}$}\hspace{.5cm}
& \multicolumn{2}{c}{\hspace{.4cm}\dda~$\leqslant \ 1 \, ~ \rm Mpc$}\hspace{.4cm}
& \multicolumn{3}{c}{\hspace{.4cm}\dda~$\leqslant \ 500 \, ~ \rm kpc$}\hspace{.5cm}
\\
\multicolumn{11}{c}{\vspace{-0.6cm}}
\\
\multicolumn{2}{c}{\rule{1.75cm}{.03cm}}
& \multicolumn{3}{c}{}
& \colhead{}
& \multicolumn{2}{c}{\rule{2.5cm}{.03cm}}
& \multicolumn{3}{c}{\rule{4cm}{.03cm}}
\\
\multicolumn{11}{c}{\vspace{-0.5cm}}
\\
\multicolumn{2}{c}{}
& \multicolumn{3}{c}{}
& \colhead{}
& \colhead{$z$}
& \colhead{Lifetime}
& \colhead{$z$}
& \colhead{Lifetime}
& \colhead{$\langle\rm SFR\rangle$}
\\
\multicolumn{11}{c}{\vspace{-0.9cm}}
\\
\multicolumn{2}{c}{}
& \multicolumn{3}{c}{\rule{6cm}{.03cm}}
& \multicolumn{6}{c}{}
\\
\colhead{}
& \colhead{}
& \multicolumn{3}{c}{$\log(M_\odot)$}
& \colhead{$\log(M_\odot $ ${\rm yr}^{-1})$}
& \colhead{}
%The unit is Gyr - no plural. We may say "10 centimeters" but write 10 cm. Same here.
& \colhead{(Gyr)}
& \colhead{}
& \colhead{(Gyr)}
& \colhead{$\log(M_\odot $ ${\rm yr}^{-1})$}
\\
\colhead{(1)}
& \colhead{(2)}
& \colhead{(3)}
& \colhead{(4)}
& \colhead{(5)}
& \colhead{(6)}
& \colhead{(7)}
& \colhead{(8)}
& \colhead{(9)}
& \colhead{(10)}
& \colhead{(11)}
}
\startdata
115  & Total & 10.95 & 9.41 & 11.91 &   -2.48 &   0.91 &  7.20 & 0.41 & 4.23 & \psp 0.02 \\
     &     A & 10.84 & 8.44 & 11.90 & \ldots  &        &       & & &\\
     &     B & 10.10 & 9.10 &  \osp9.53 & \ldots  &        &       & & &\\
     &     C &  \osp9.94 & 9.00 &  \osp7.86 &   -2.48 &        &       & & &\\
377  & Total & 11.12 & 9.89 & 11.93 &   -1.15 &   1.17 &  8.25 & 0.69 & 6.10 &  \psp0.36 \\
     &     A & 10.94 & 9.08 & 11.93 & \ldots  &        &       & & &\\
     &     B & 10.29 & 9.28 &  \osp7.86 &   -2.30 &        &       & & &\\
     &     C & 10.04 & 9.30 &  \osp6.37 &   -1.24 &        &       & & &\\
     &     D &  \osp9.92 & 9.39 &  \osp9.41 &   -2.13 &        &       & & &\\
     &     E &  \osp9.80 & 8.34 &  \osp7.48 & \ldots  &        &       & & &\\
518  & Total &  \osp9.23 & 9.49 &  \osp8.78 &   -1.18 &   3.06 & 11.41 & 1.39 & 8.90 & -0.44 \\
     &     A &  \osp8.82 & 9.25 &  \osp7.86 & \ldots  &        &       & & &\\
     &     B &  \osp8.79 & 8.93 &  \osp8.56 &   -1.63 &        &       & & &\\
     &     C &  \osp8.61 & 8.69 &  \osp8.21 &   -1.36 &        &       & & &\\
1056 & Total & 10.78 & 9.57 & 11.50 &   -0.95 &   0.62 &  5.73 & 0.09 & 1.13 & -0.21 \\
     &     A & 10.64 & 8.19 & 11.50 & \ldots  &        &       & & &\\
     &     B & 10.16 & 9.04 &  \osp7.86 &   -1.64 &        &       & & &\\
     &     C &  \osp9.27 & 9.39 &  \osp9.08 &   -1.05 &        &       & & &\\
1119 & Total & 10.71 & 9.84 & 11.27 & \psp0.16 &   0.99 &  7.60 & 0.41 & 4.23 &  \psp0.23 \\
     &     A & 10.47 & 9.12 & 11.25 & \ldots  &        &       & & &\\
     &     B & 10.13 & 9.45 &  \osp7.86 &   -1.79 &        &       & & &\\
     &     C &  \osp9.93 & 9.44 &  \osp9.94 & \psp0.16&        &       & & &\\
1441 & Total & 10.76 & 9.83 & 11.51 &   -1.13 &   0.99 &  7.60 & 0.36 & 3.86 & -0.63 \\
     &     A & 10.59 & 9.42 & 10.09 &   -2.54 &        &       & & &\\
     &     B & 10.21 & 9.53 & 11.49 &   -1.15 &        &       & & &\\
     &     C &  \osp9.36 & 8.90 &  \osp7.86 &   -3.86 &        &       & & &\\
1598 & Total & 11.01 & 9.67 & 11.92 &   -0.16 &   0.62 &  5.73 & 0.17 & 2.09 &  \psp0.51 \\
     &     A & 10.87 & 8.48 & 11.92 & \ldots  &        &       & & &\\
     &     B & 10.15 & 9.36 &  \osp8.22 &   -0.16 &        &       & & &\\
     &     C & 10.10 & 9.25 &  \osp7.86 &   -3.20 &        &       & & &\\
     &     D &  \osp9.40 & 8.51 &  \osp8.06 & \ldots  &        &       & & &\\
1757 & Total & 10.69 & 9.06 & 11.29 & \psp0.62&   1.17 &  8.25 & 0.28 & 3.13 & -0.27 \\
     &     A & 10.52 & 8.36 & 11.29 & \psp0.61&        &       & & &\\
     &     B & 10.13 & 8.28 &  \osp6.38 &   -2.04 &        &       & & &\\
     &     C &  \osp9.24 & 8.85 &  \osp8.83 &   -0.88 &        &       & & &\\
1773 & Total &  \osp9.89 & 9.76 & 10.85 &   -0.32 &   2.62 & 11.01 & 0.83 & 6.84 &  \psp0.41 \\
     &     A &  \osp9.77 & 9.63 & 10.85 &   -0.33 &        &       & & &\\
     &     B &  \osp8.86 & 8.94 &  \osp7.86 &   -2.21 &        &       & & &\\
     &     C &  \osp9.10 & 8.81 &  \osp7.86 &   -2.82 &        &       & & &\\
2143 & Total & 10.27 & 9.70 & 11.02 &   -0.26 &   1.77 &  9.78 & 0.36 & 3.86 &  \psp0.01 \\
     &     A & 10.17 & 9.40 & 11.01 & \ldots  &        &       & & &\\
     &     B &  \osp9.33 & 9.02 &  \osp9.01 &   -0.55 &        &       & & &\\
     &     C &  \osp9.23 & 9.14 &  \osp7.86 &   -0.58 &        &       & & &\\
\enddata
\vspace{.1cm}
\hspace{-.6cm}
\renewcommand{\baselinestretch}{0.8}\selectfont
\begin{minipage}[0.01\textheight]{1.05\textwidth}
  {
\footnotesize
     {\bf Note.} Member galaxy properties at $z=0$ for CGs identified with \clues\ in the Guo2010a milli-Millennium database \citep[sCGs;][\mmsg]{guo2011}. Column~(1):~sCG ID. Column~(2):~member galaxy ID (A to E, according to decreasing stellar mass). Column~(3):~stellar mass. Column~(4):~cold gas mass. Column~(5):~hot gas mass. Column~(6):~star formation rate. Column~(7):~redshift at which members of compact group come within \dda~$\leqslant 1$ Mpc of the most massive galaxy for the first time. Column~(8):~lifetime as a compact group from aforementioned redshift to $z=0$. Columns (9)-(10):~same as Columns (7)-(8), but instead using \dda~$\leqslant 500$ kpc. Column~(11):~mean star formation rate during lifetime within $\leqslant 500$ kpc. In Column~(6), where there are no entries, the Guo2010a semianalytical model assigns values of zero.
     }
\end{minipage}
\renewcommand{\baselinestretch}{1}\selectfont
  \vspace{-0.8cm}
\end{deluxetable*}

\subsection{Detections with the \clues\ 
Algorithm}\label{sec-clues}

The CLUstEring with local Shrinking (\clues)\footnote{\href{http://CRAN.R-project.org/package=clues}{http://CRAN.
R-project.org/package=clues}}package \citep{wang2007,chang2010} is part of the R statistical software environment \citep{rteam2012}. 
It is a cluster\footnote{In this Section the term ``cluster" refers to structures within a parameter space of data and should not be confused with the term commonly used in astronomy.} analysis algorithm aiming to make minimal assumptions about datasets and an optimal identification of cluster numbers. \clues\ is thus an unsupervised clustering analysis algorithm, where in particular the number of clusters is not an input value provided by the user. This is a major advantage for the type of analysis performed in this paper, where the number of clusters is not known in advance.

Briefly, \clues\ identifies cluster centers by effectively drawing data points together (``shrinking'') until they converge on the cluster center, similarly to the behavior of point-mass particles in a gravitational field. In each iteration, the median position of all points in the smallest sphere containing $K$ nearest neighbors replaces all previous positions in the sphere until the distance $d$ between points in successive iterations becomes smaller than a stopping criterion $\epsilon$. Individual data points are then assigned to specific clusters (a process called ``partitioning'') by successively calculating pairwise distances between neighboring points and identifying outliers in the distance distribution. The number $K$ is also iteratively optimized. Starting with a small $K$,  its value is increased until one of two well-established indices of cluster strength are optimized, namely, Calinski and Harabsz (CH; \citealt{CHscore}) or Silhouette \citep{silhouette}. 

For this paper, we used the \clues\ (2.15.2) implementation in R to detect galaxy concentrations similar to CGs in the \mms. The data points searched by \clues\ were simply the 3D Euclidean coordinates of galaxies in the catalog. We restricted the search to $z=0$ and $M_V<-18$ to select against dwarf galaxies and mimic the \citet{hickson1982} compactness criterion. As the Silhouette index has proven more successful in identifying clusters with irregular shapes, we used this index as a measure of cluster strength. After completing the \clues\ analysis, for each cluster we identified the minimum galaxy magnitude, $M_{V,{\rm min}}$, and then removed any galaxies with magnitudes $\geqslant M_{V,{\rm min}} + 3$, mimicking the \citet{hickson1982} magnitude range. We calculated a radius, defined as the maximum distance between the cluster center 
%\todon{I'm avoiding medoid} 
and each member galaxy, as well as the separation from all other clusters. We discarded clusters that were separated from their nearest neighbor by less than three times their diameter, aiming to mimic in 3D the \citet{hickson1982} angular isolation criterion.  Finally, clusters that had fewer than three galaxy members were also discarded, to match the standard observational practice for CGs.

The result of this process was a list of 12 galaxy groups, containing 41 galaxies at $z=0$ in \mmsg. Although the \clues-based selection is in 3D space and thus not fully consistent with the selection of observed CGs, the identified systems are still isolated, compact galaxy concentrations, and we refer to them as simulated compact groups (sCGs). For two of these there were unphysical discontinuities in positional information along the merger trees, likely due to proximity to the edge of the simulation box. We could thus not reliably construct merger trees and discarded these systems.

Among the remaining 10 sCGs (\tr{tab-simdata}), one had four members (sCG~1598), and one had five members (sCG~377). The remaining eight had three members.

\vspace{.2cm}
\subsection{Galaxy Properties and Visualization}
\vspace{.2cm}
\renewcommand{\baselinestretch}{0.6}\selectfont
\begin{figure*}[ht]
  \centering
  \includegraphics[scale=.42]{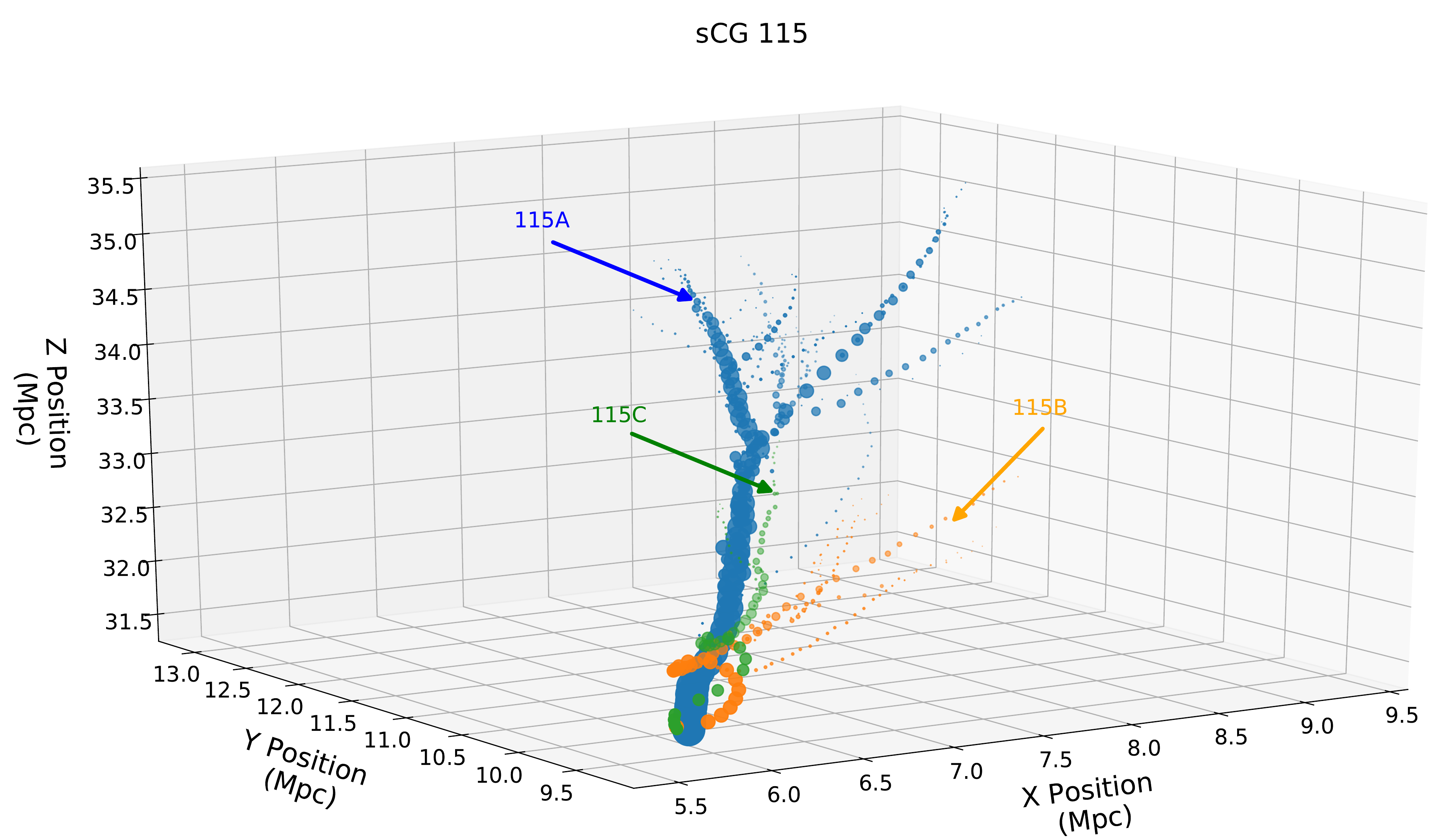}
  \caption{3D merger tree for sCG~115, reconstructed from \mmsg\ data. Member galaxies are shown as filled colored circles, and circle sizes scale with \mstar. Galaxies evolve from low to high \mstar, and the final configuration of the group at $z=0$ is near the lower center of the figure as viewed in perspective. Galaxies are labeled A to C in order of decreasing \mstar\ at $z=0$.} 
  \label{fig-sCG115_3D}
\end{figure*}
\renewcommand{\baselinestretch}{1}\selectfont
We use the ID numbers assigned to identified sCGs by the \clues\ algorithm. Strictly, these only apply to sCGs at $z=0$; however, for simplicity, we use them to refer to sCGs, as well as their past constituents, throughout their evolution over cosmic time along their merger trees \citep{lee2014}.  The merger tree information is extracted from the \mmsg\ database using the $z=0$ \mmsg\ galaxy IDs (preserved by \clues).  The \mmsg\ catalog contains information for 82 galaxy properties over the 64 redshift snapshots of the simulation. Apart from the redshift of each snapshot, we used SQL catalog queries 
to obtain information for 10 properties in each snapshot, namely, stellar mass (\mstar), cold gas mass (\mcgas), hot gas mass (\mhgas), SFR, and $xyz$ positions and velocities. 

\renewcommand{\baselinestretch}{0.6}\selectfont
\begin{figure*}[ht]
  %\vspace{-1cm}
   \centering
    \begin{minipage}{.40\linewidth}
        \centering
        \subfigure{\includegraphics[scale=.562]{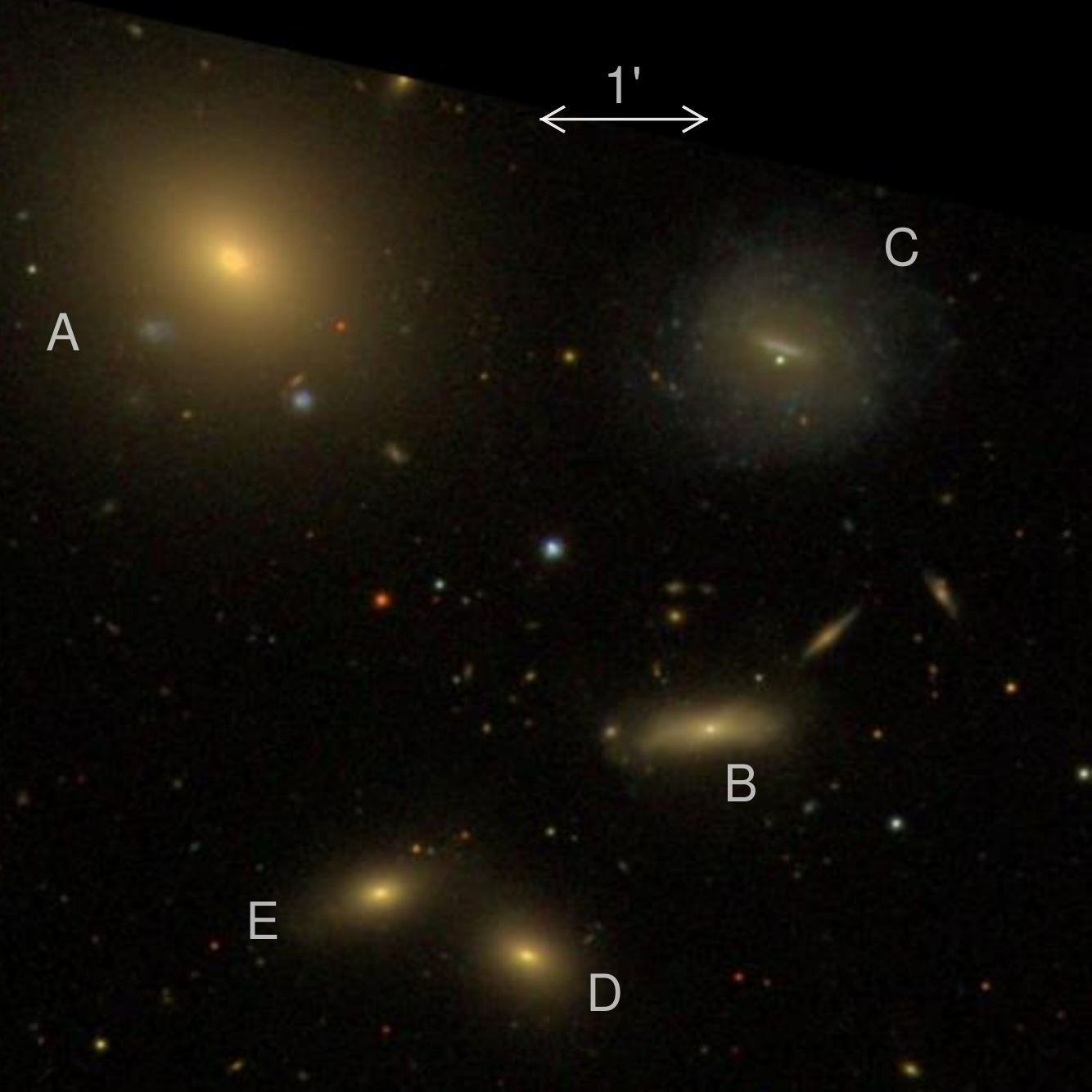}}\\[-.5ex]
        \subfigure{\includegraphics[scale=.39]{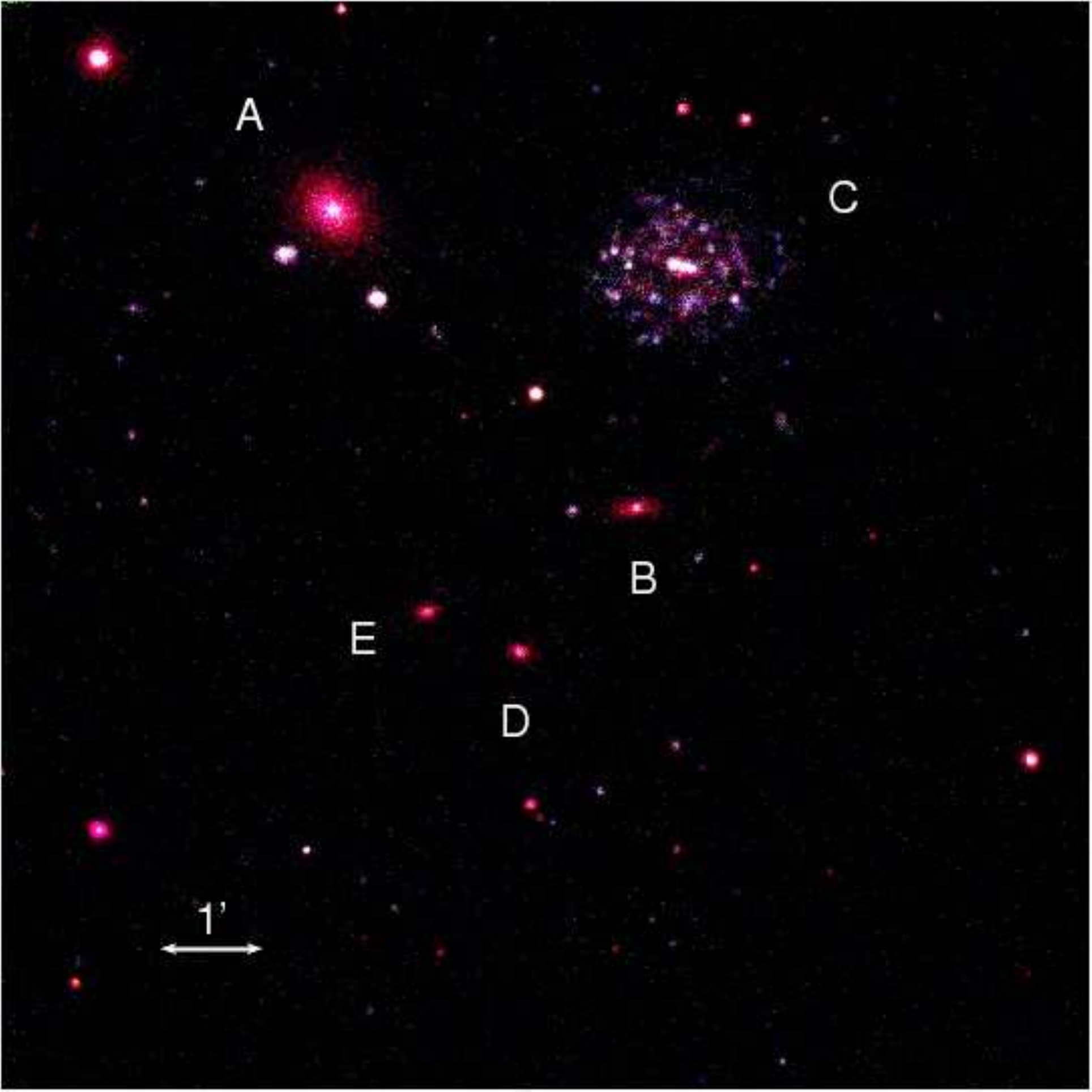}}
        \end{minipage}%ccccc
    \begin{minipage}{.595\linewidth}
        \centering
        \subfigure{\includegraphics[scale=.45]{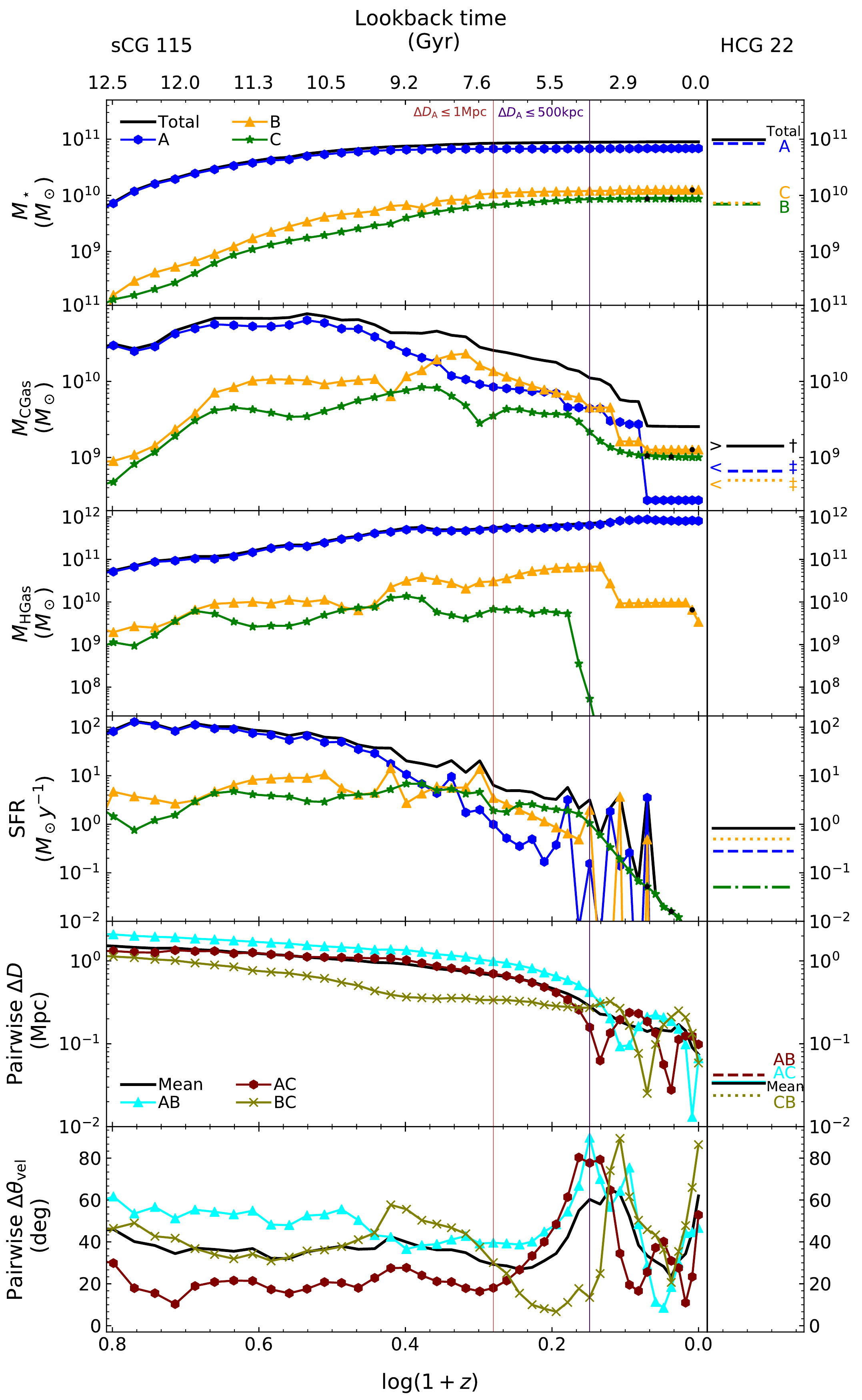}}
    \end{minipage}
   \caption{\footnotesize Top Left: SDSS $u,g,r,i,z$ optical image of
     HCG~22 (all SDSS images in this paper from \href{https://skyserver.sdss.org/dr14/en/home.aspx}{ https://skyserver.sdss.org/dr14/en/home.aspx}).
     Bottom Left:
     \swift/\uvot\ three-band image of HCG~22, with blue, green, and
     red colors corresponding to the \wtwo, \mtwo, and \wone\ filters,
     respectively \citep[][and/or \swift\ archive for this and all
       \swift/UVOT images in this paper]{tzanavaris2010}. Galaxies D
     and E are background objects. Right: from top to bottom,
     left hand panels show cosmic evolution of \mstar, cold gas mass,
     hot gas mass, SFR, pairwise galaxy separation, and pairwise
     velocity vector angle for sCG~115. The top axis gives the
     corresponding look-back time, and the bottom axis gives the
     redshift. The color-coding for each galaxy of sCG~115 corresponds
     directly to that in the 3D merger tree shown in
     \fr{fig-sCG115_3D}. In each panel, the solid black line shows
     either the combined group total (for \mstar, \mcgas, \mhgas, and
     SFR) or the mean (for pairwise $\Delta D$ and $\Delta \theta_{\rm
       vel}$). The brown vertical line across all panels indicates
     when CG members come within $\leqslant 1$ Mpc of the most
     massive galaxy for the first time, while the purple vertical line
     shows the same for $\leqslant 500$ kpc. In the galaxy separation panel,
     AB gives the separation between galaxies A and B, AC between
     galaxies A and C, etc. The color-coding and symbol coding for this
     panel also apply to the velocity vector angle panel. The black
     points indicate interpolated data points where data from the
     simulation were missing. Right hand panels show values for the
     same quantities as in the left hand panels but for HCG~22
     galaxies at $z=0$ (\tr{tab-hcgdata}), where available. In this
     case the HCG~22B and HCG~22C galaxies are not in order of decreasing
     stellar mass; the line color/style still reflects decreasing
     stellar mass, so that sCG~115B is matched with HCG~22C and
     sCG~115C with HCG~22B. The galaxy separations for HCGs are 2D
     projections and will therefore be underestimates.}
    \label{fig-sCG115_vs_HCG22}
\end{figure*}

\renewcommand{\baselinestretch}{1}\selectfont

\begin{figure*}[ht]
  \centering
  \includegraphics[scale=.42]{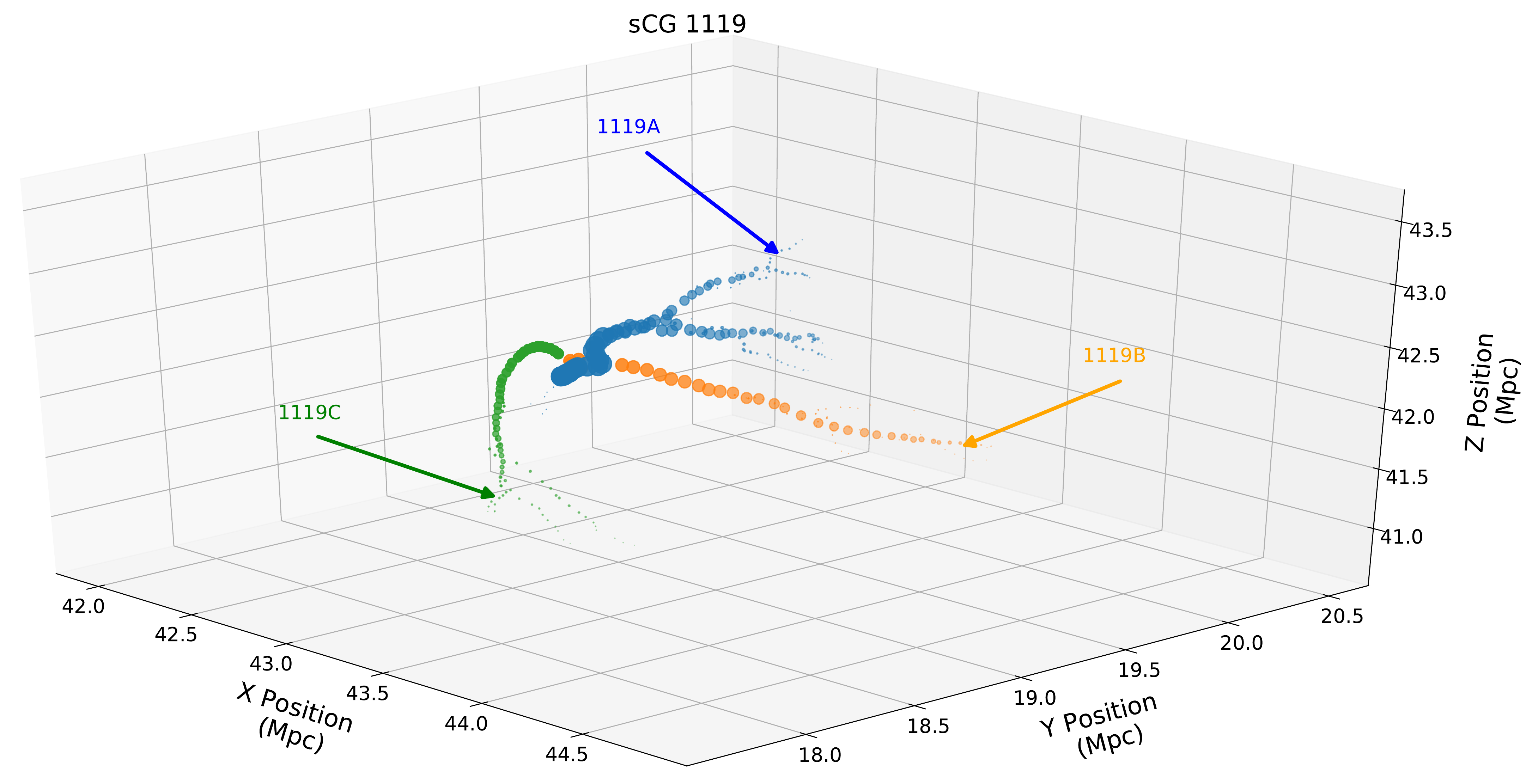}
  \caption{Same as \fr{fig-sCG115_3D}, but now showing sCG~1119. The final configuration of the group at $z=0$ is roughly at the center left of the figure as viewed in perspective. Galaxies are labeled A to C in order of decreasing \mstar\ at $z=0$.} 
  \label{fig-sCG1119_3D}
\end{figure*}

\begin{figure*}[ht]
  \centering
    \begin{minipage}{.40\linewidth}
        \centering
        \subfigure{\includegraphics[scale=.562]{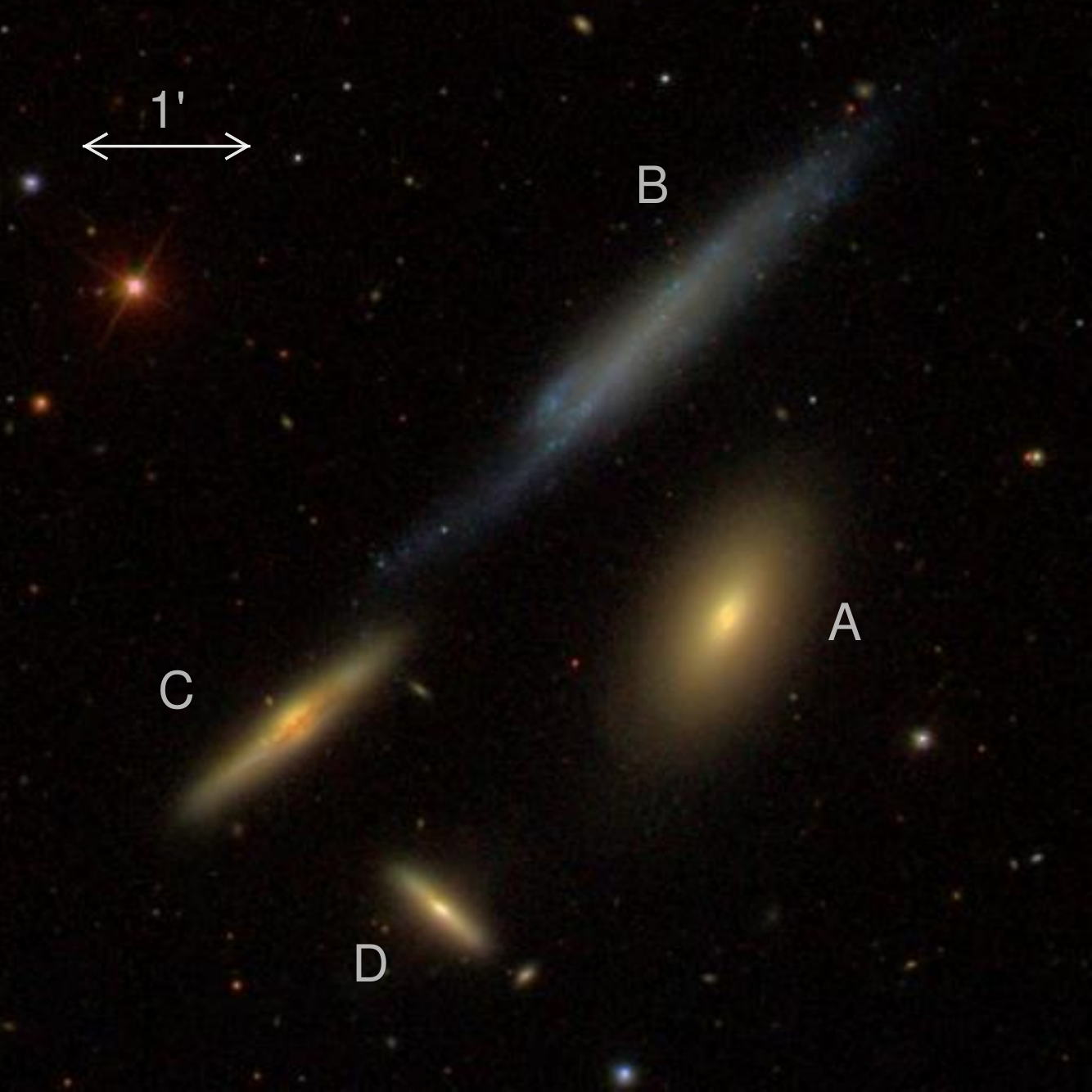}}\\[-.5ex]
        \subfigure{\includegraphics[scale=.39]{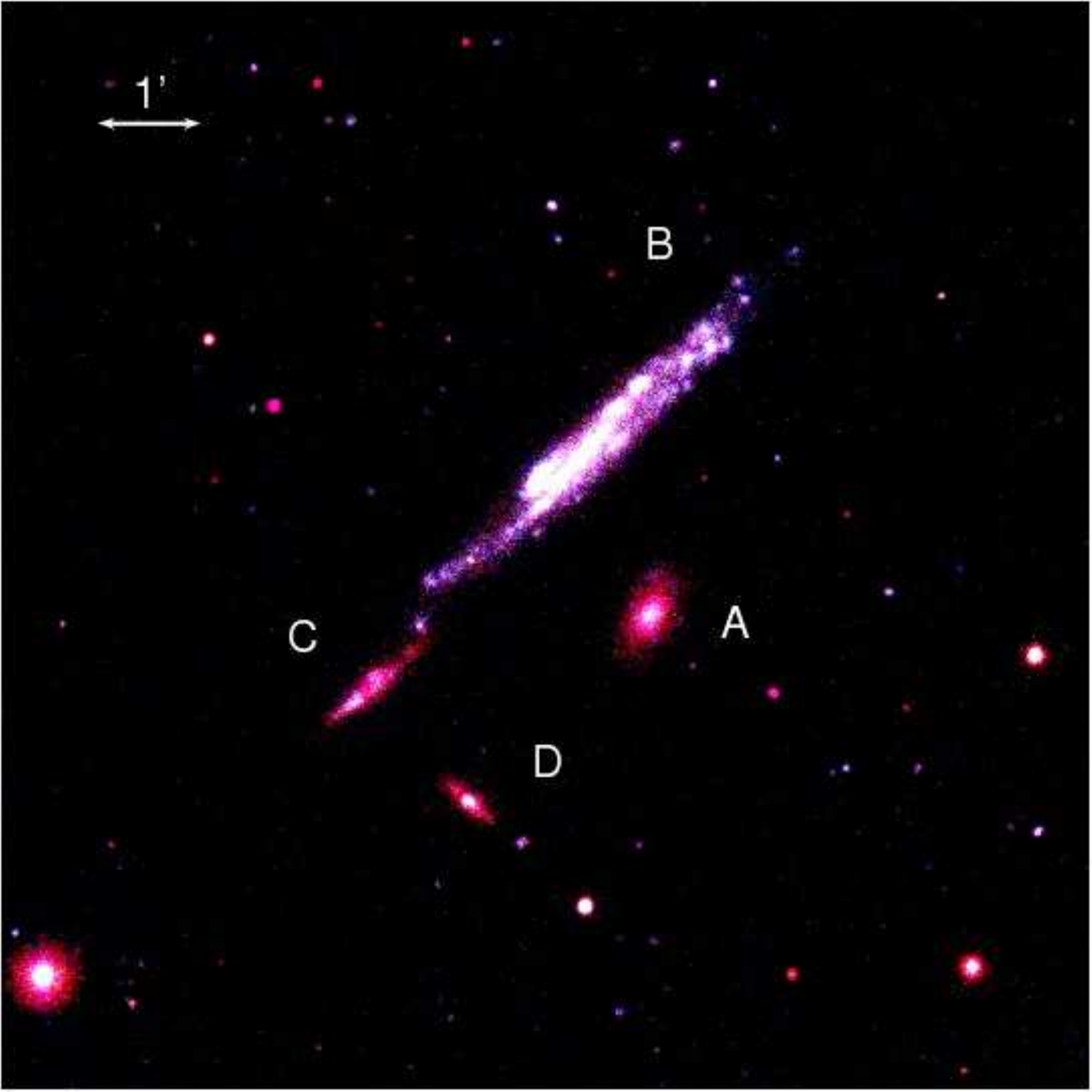}}
        \end{minipage}%
    \begin{minipage}{.595\linewidth}
        \centering
        \subfigure{\includegraphics[scale=.45]{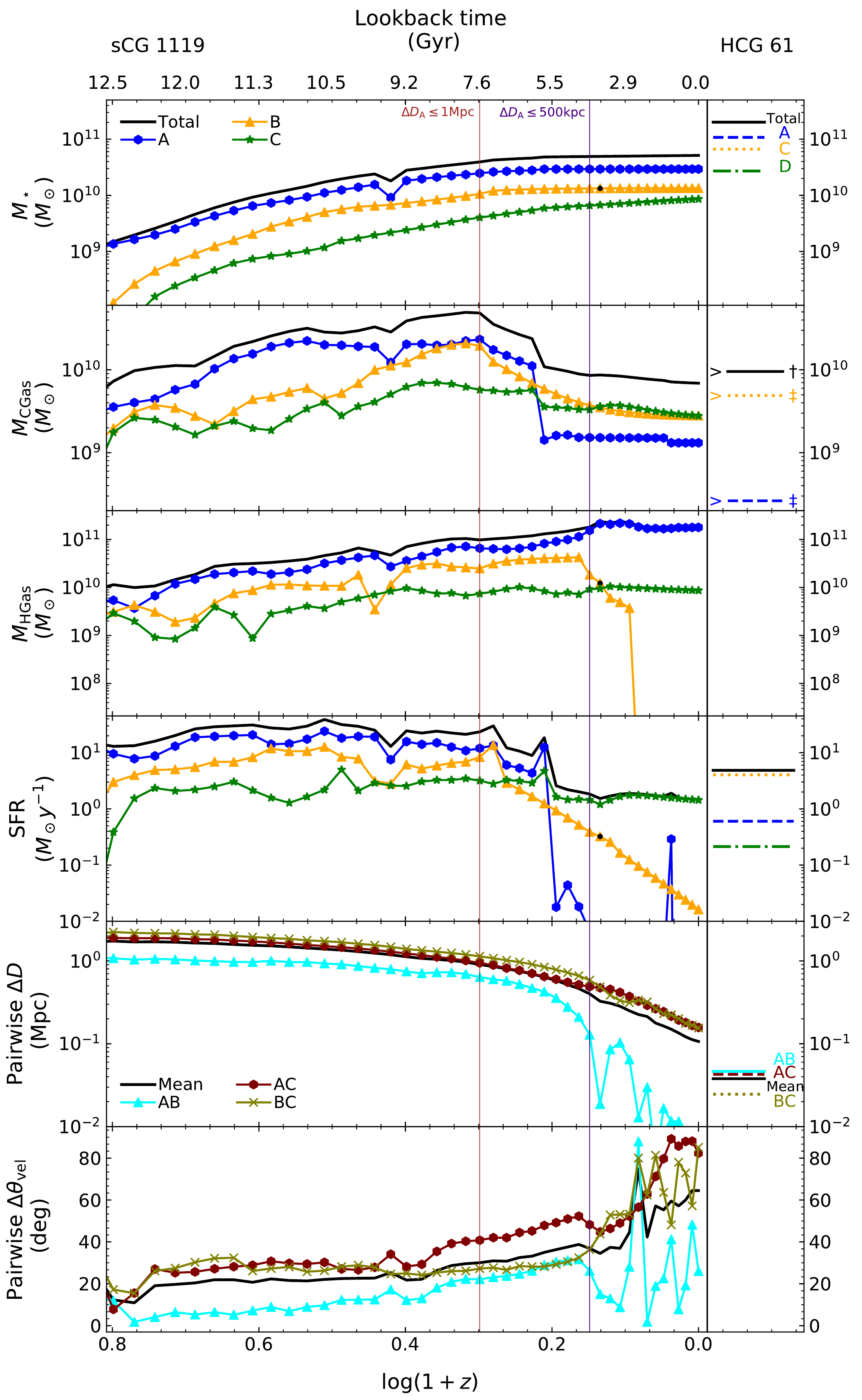}}
    \end{minipage}
   \caption{\footnotesize Top Left: SDSS optical image as in \fr{fig-sCG115_vs_HCG22}, but now showing HCG~61. Bottom Left: \swift/\uvot\ color image as in \fr{fig-sCG115_vs_HCG22}, also for HCG~61. Galaxy B is a foreground object. Right: Same as \fr{fig-sCG115_vs_HCG22}, but instead showing sCG~1119 vs HCG~61. The color-coding for each sCG~1119 galaxy corresponds directly to that in the 3D merger tree shown in \fr{fig-sCG1119_3D}. In order of decreasing stellar mass, sCG~1119B is best matched to HCG~61C, and sCG~1119C to HCG~61D.}
  \label{fig-sCG1119_vs_HCG61}
\end{figure*}

\begin{figure*}[ht]
  \centering
  \includegraphics[scale=.42]{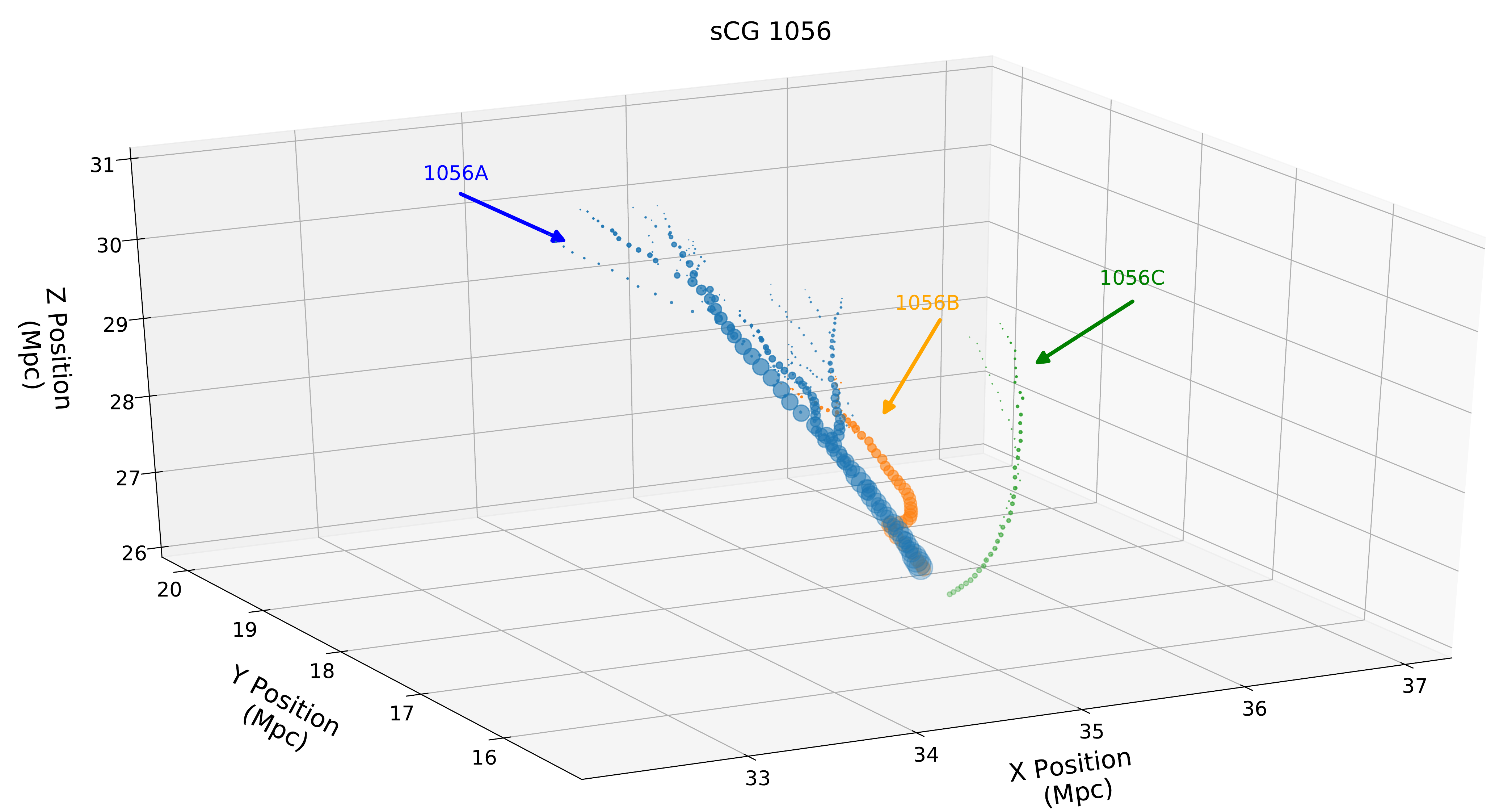}
  \caption{Same as \fr{fig-sCG115_3D}, but now showing sCG~1056. The final configuration of the group at $z=0$ is roughly at the lower center right of the figure as viewed in perspective. Galaxies are labeled A to C in order of decreasing \mstar\ at $z=0$.} 
  \label{fig-sCG1056_3D}
\end{figure*}

\begin{figure*}[ht]
  \centering
    \begin{minipage}{.40\linewidth}
        \centering
        \subfigure{\includegraphics[scale=.362]{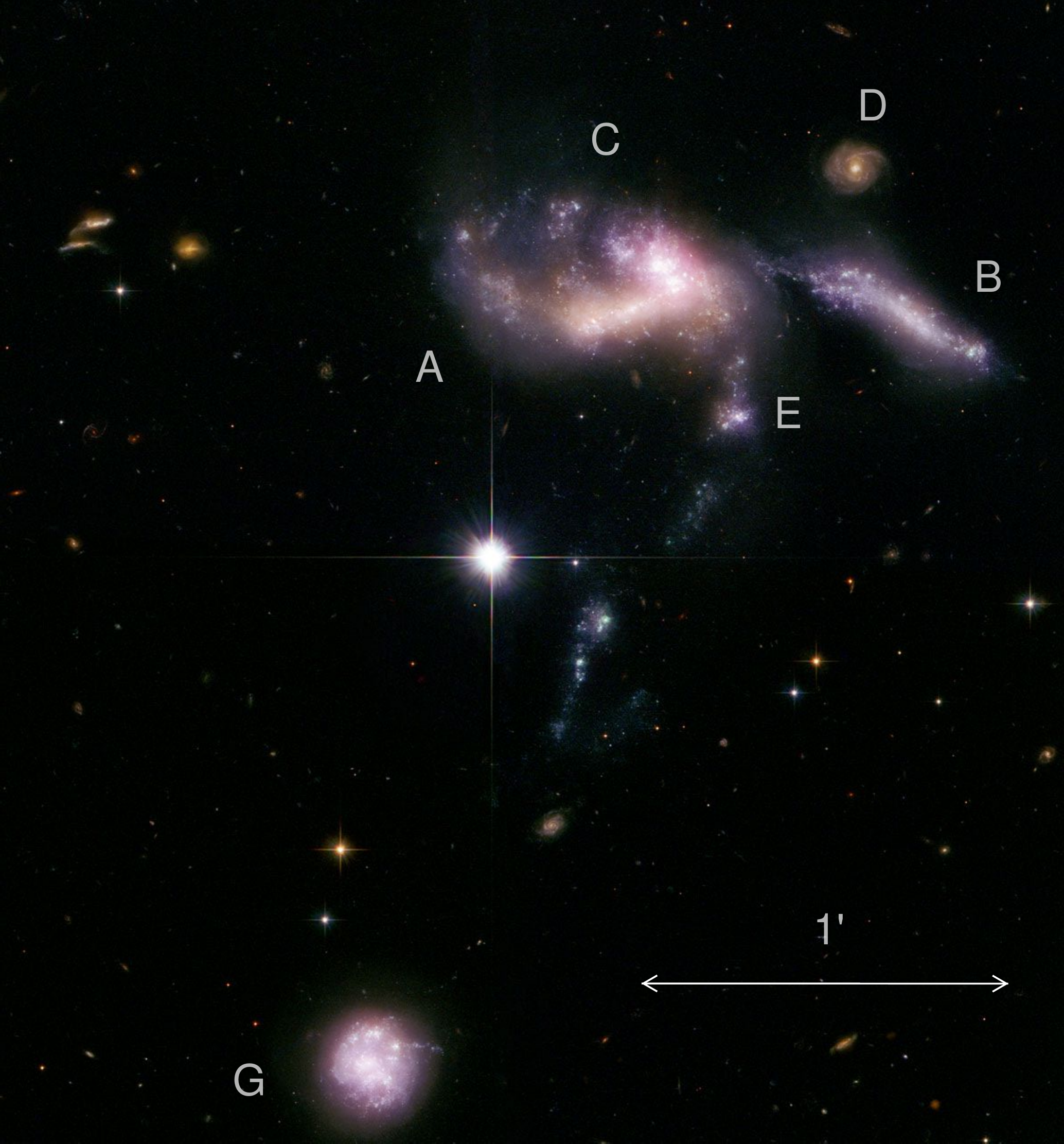}}\\[-.5ex]
        \subfigure{\includegraphics[scale=.39]{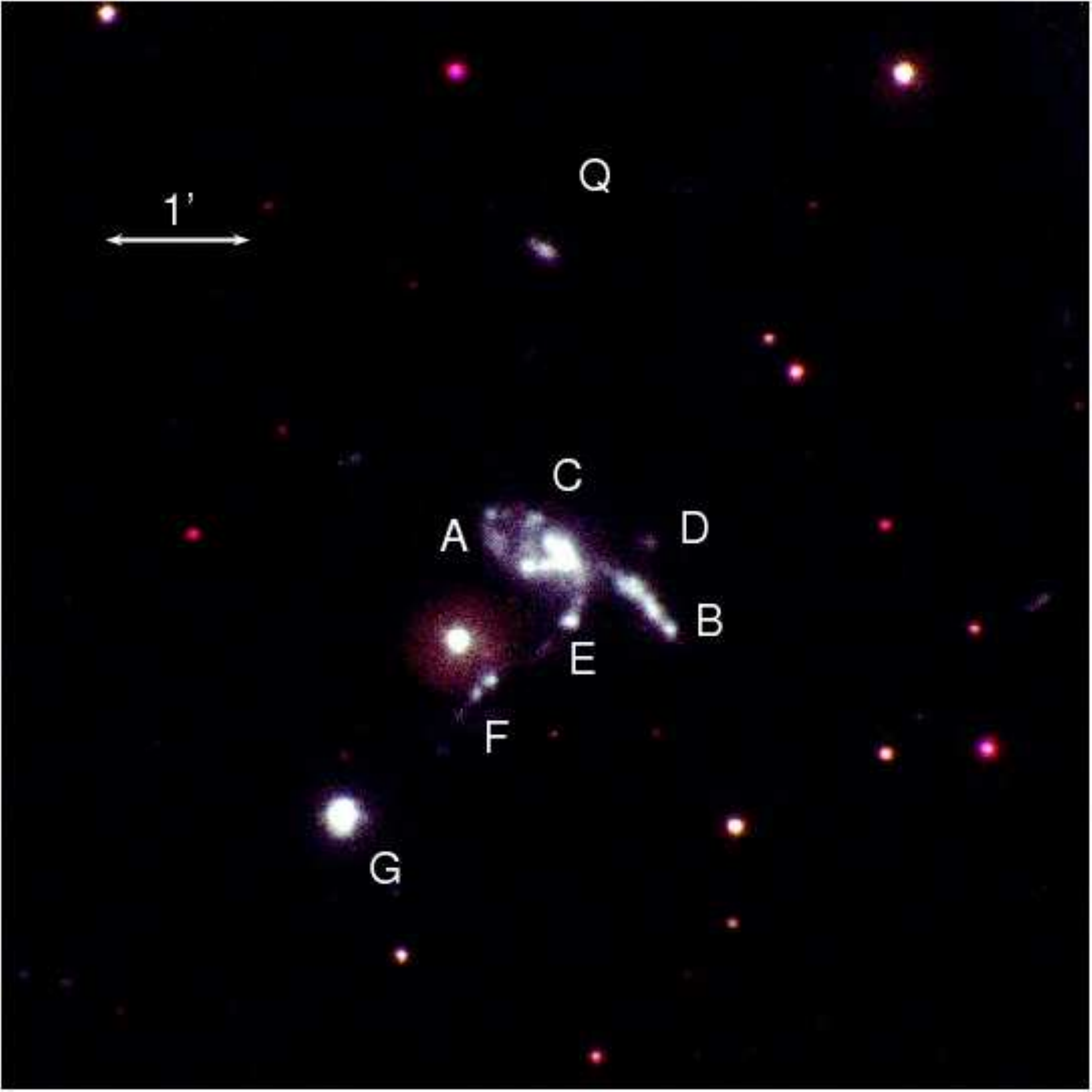}}
        \end{minipage}%
    \begin{minipage}{.595\linewidth}
        \centering
        \subfigure{\includegraphics[scale=.45]{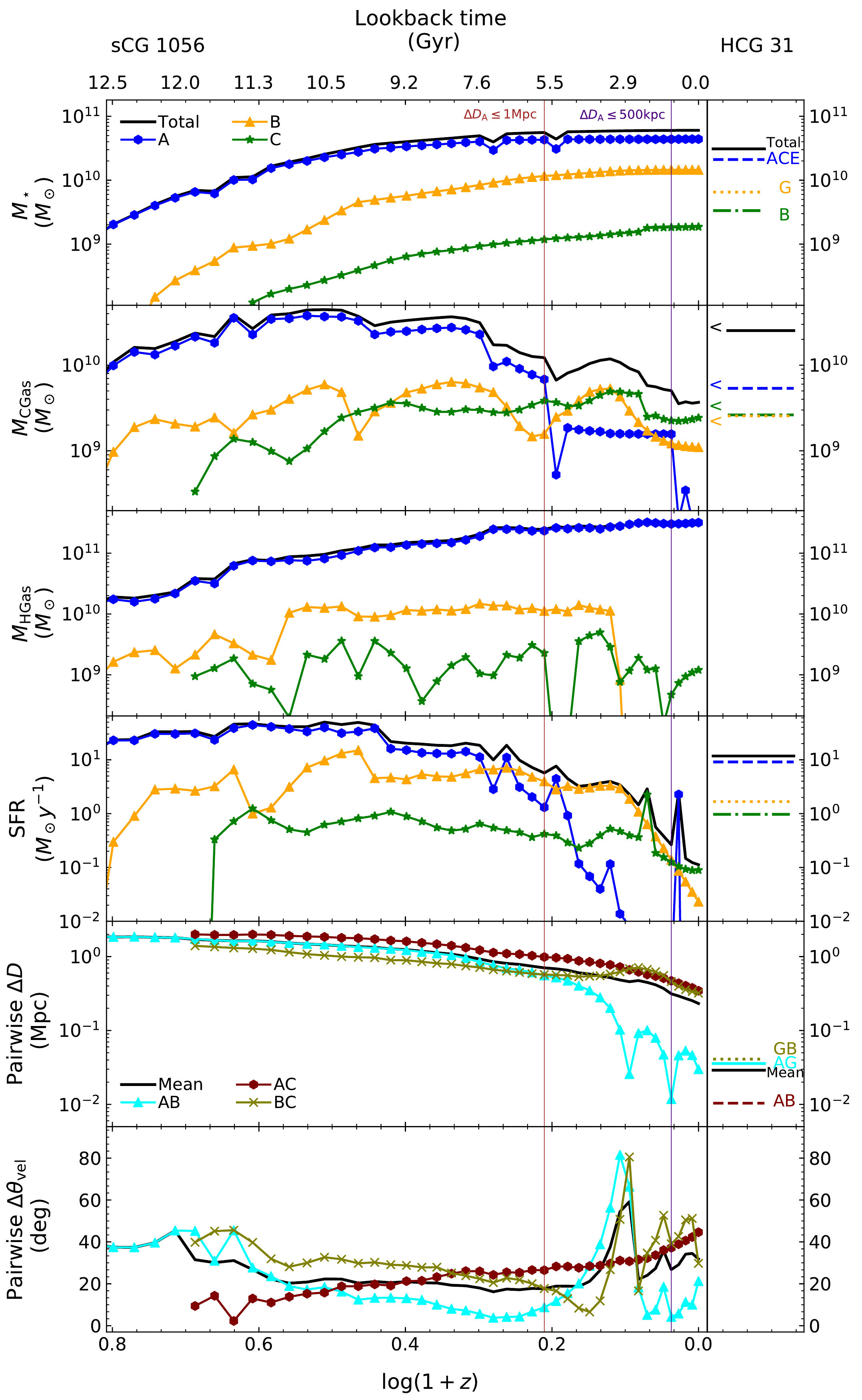}}
    \end{minipage}
     \caption{Top Left: Composite image of HCG~31 combining \hst\ ACS ($B_{435}$, $V_{606}$, and $I_{814}$), \hst\ WFPC2 ($B_{439}$, $V_{555}$, and $I_{814}$), \galex\ (NUV), and \spitzer\ (8\mic\ NIR). Image credit: NASA, ESA, S. Gallagher and J. English(\href{http://www.spacetelescope.org/images/opo1008a/}{http://www.spacetelescope.org/images/opo1008a/}). Bottom Left:  \swift/\uvot\ color image as in \fr{fig-sCG115_vs_HCG22}, but now showing HCG~31. Galaxy D is a background object, Q is likely physically associated but is quite small compared to the other associated galaxies \citep{rubin1990}, and F is a tidal feature \citep{gallagher2010}. Galaxies A and C appear to be undergoing an ongoing merger, and E is likely a tidal feature \citep{lopez-sanchez2004}(Galaxy E may be a tidal remnant and not a distinct galaxy.). Right: Same as \fr{fig-sCG115_vs_HCG22}, but instead showing sCG~1056 vs HCG~31. The color-coding for each sCG~1056 galaxy corresponds directly to that in the 3D merger tree shown in \fr{fig-sCG1056_3D}. In order of decreasing stellar mass, sCG~1056A is best matched to HCG~31ACE, sCG~1056B to HCG~31G, and sCG~1056C to HCG~31B. In the pairwise galaxy separation and velocity vector angle panels of HCG~31, we use the location of galaxy A to define the center of ACE.}
\label{fig-sCG1056_vs_HCG31}
\end{figure*}

\begin{figure*}[ht]
  \centering
  \includegraphics[scale=.42]{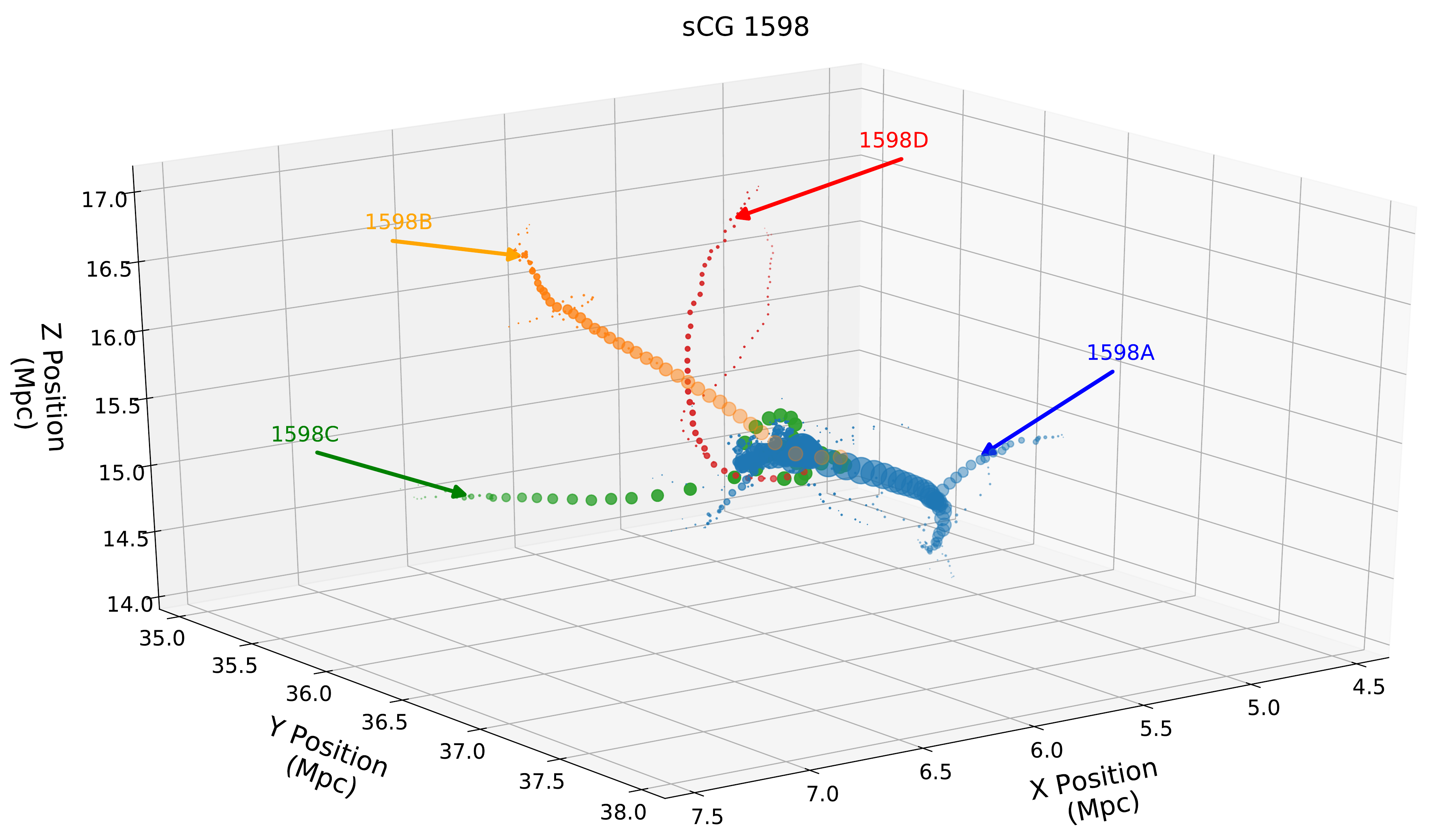}
  \caption{Same as \fr{fig-sCG115_3D}, but now showing sCG~1598. The final configuration of the group at $z=0$ is roughly at the center of the figure as viewed in perspective. Galaxies are labeled A to D in order of decreasing \mstar\ at $z=0$.}
  \label{fig-sCG1598_3D}
\end{figure*}

\begin{figure*}[ht]
  \centering
    \begin{minipage}{.35\linewidth}
      \vspace{-2cm}
        \centering
        \subfigure{\includegraphics[scale=.20]{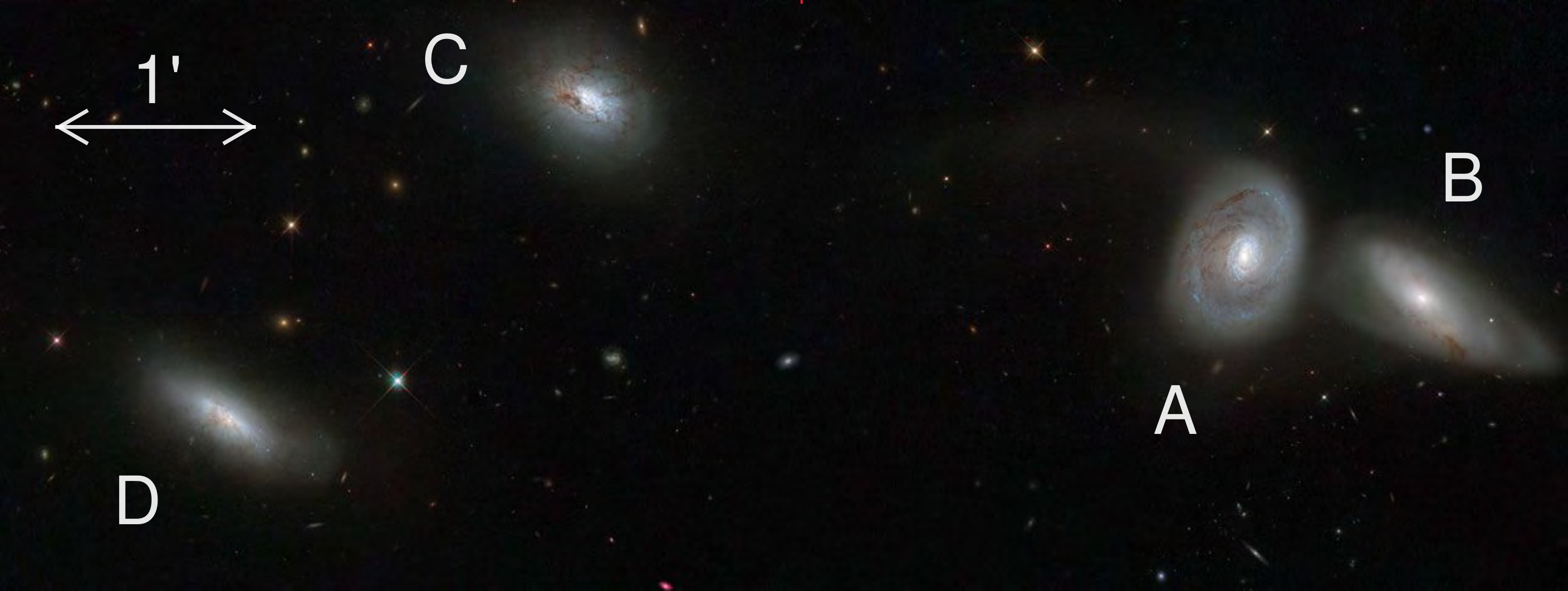}}\\[-.7ex]
        \subfigure{\includegraphics[scale=.305]{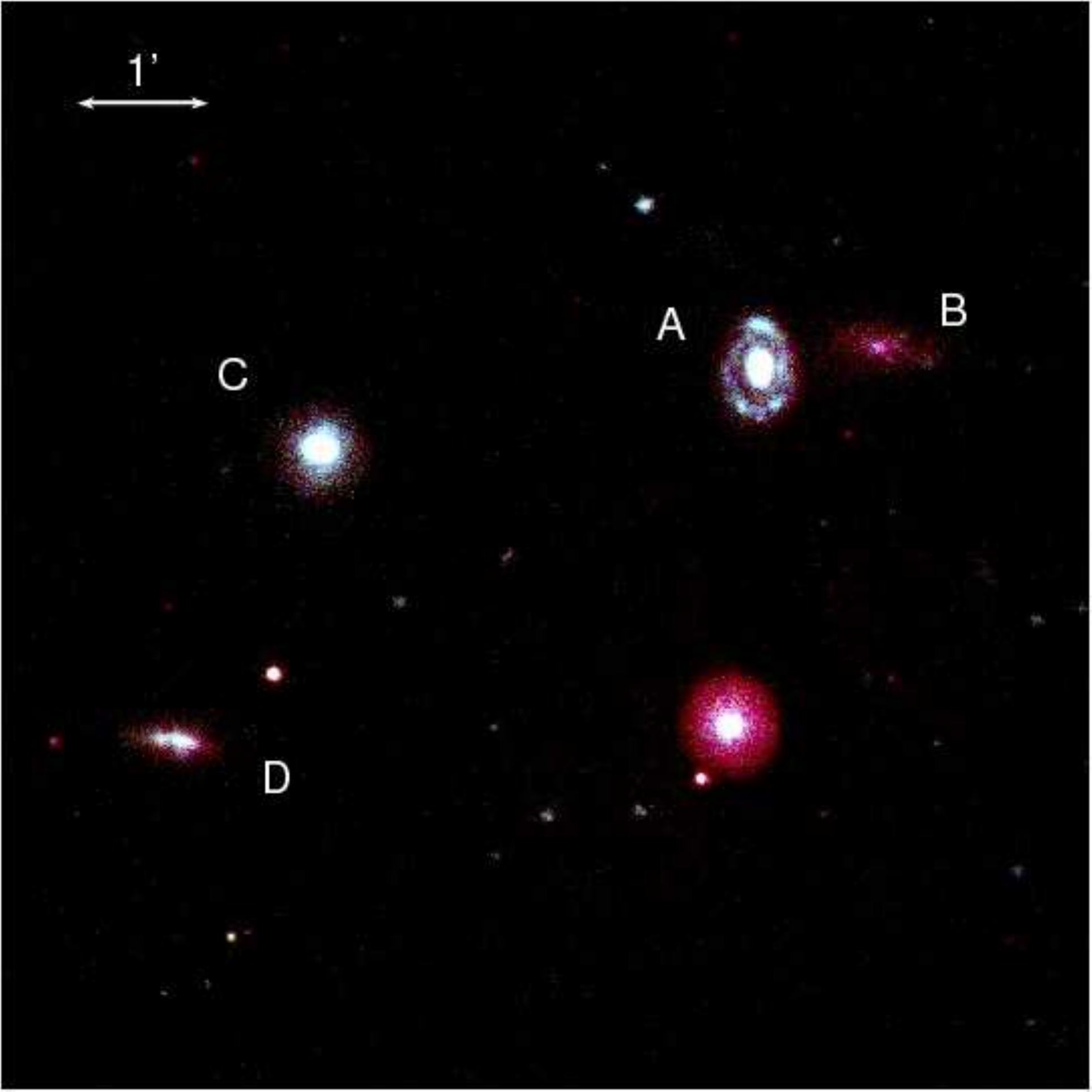}}\\[-.7ex]
        \subfigure{\includegraphics[scale=.227]{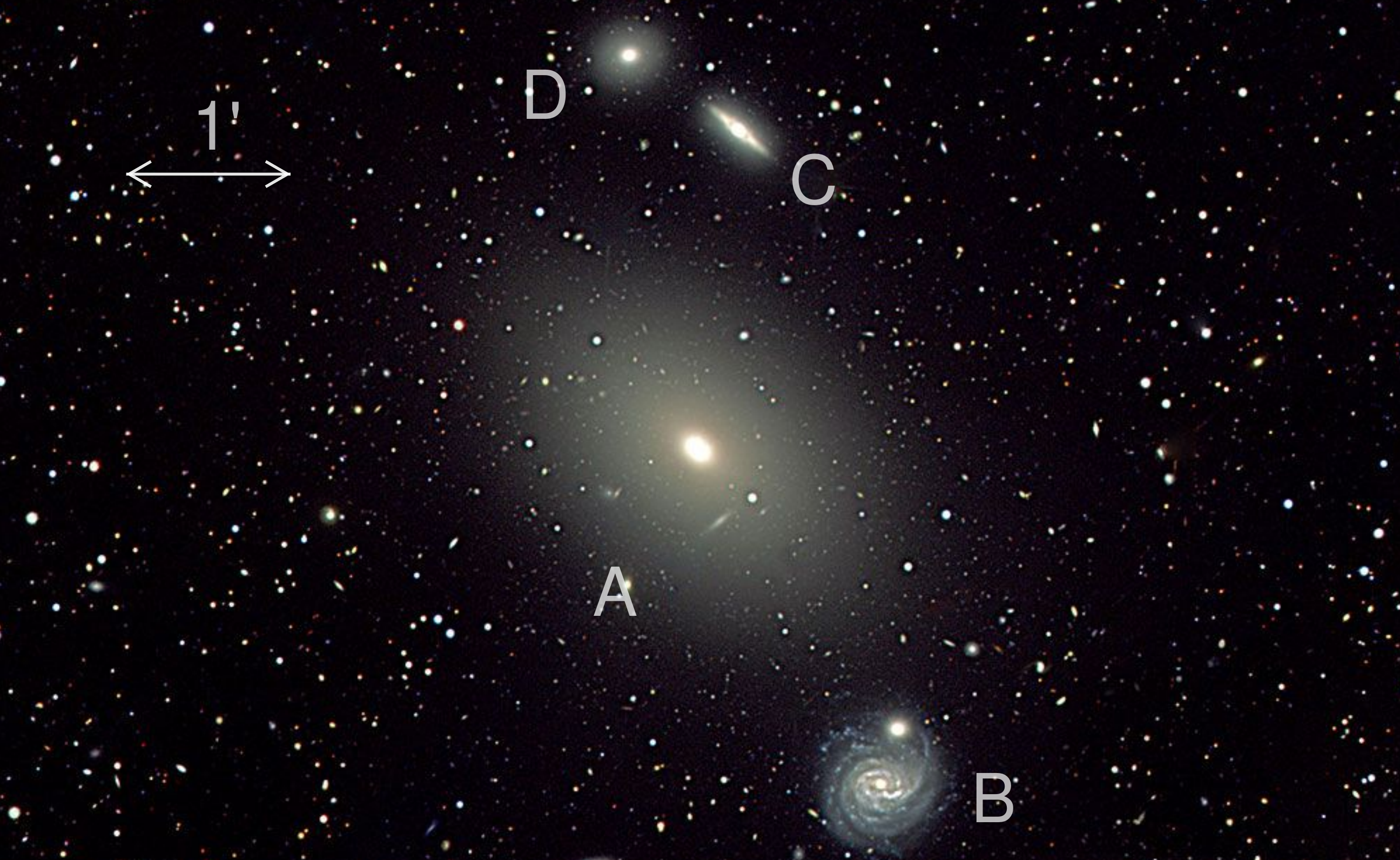}}\\[-.7ex]
        \subfigure{\includegraphics[scale=.305]{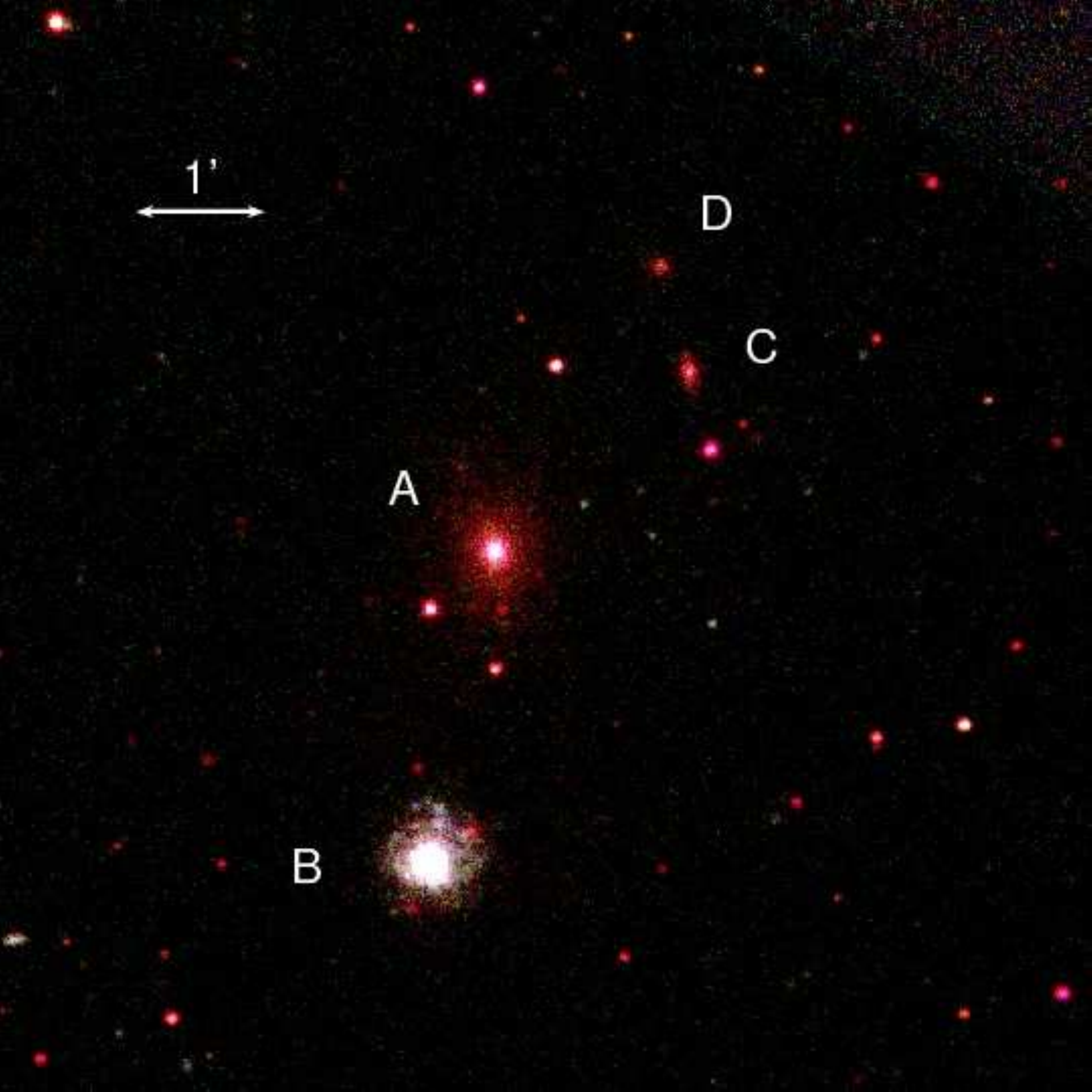}}
    \end{minipage}%
    \begin{minipage}{.62\linewidth}
    \vspace{-1cm}
        \centering
        \subfigure{\includegraphics[scale=.55]{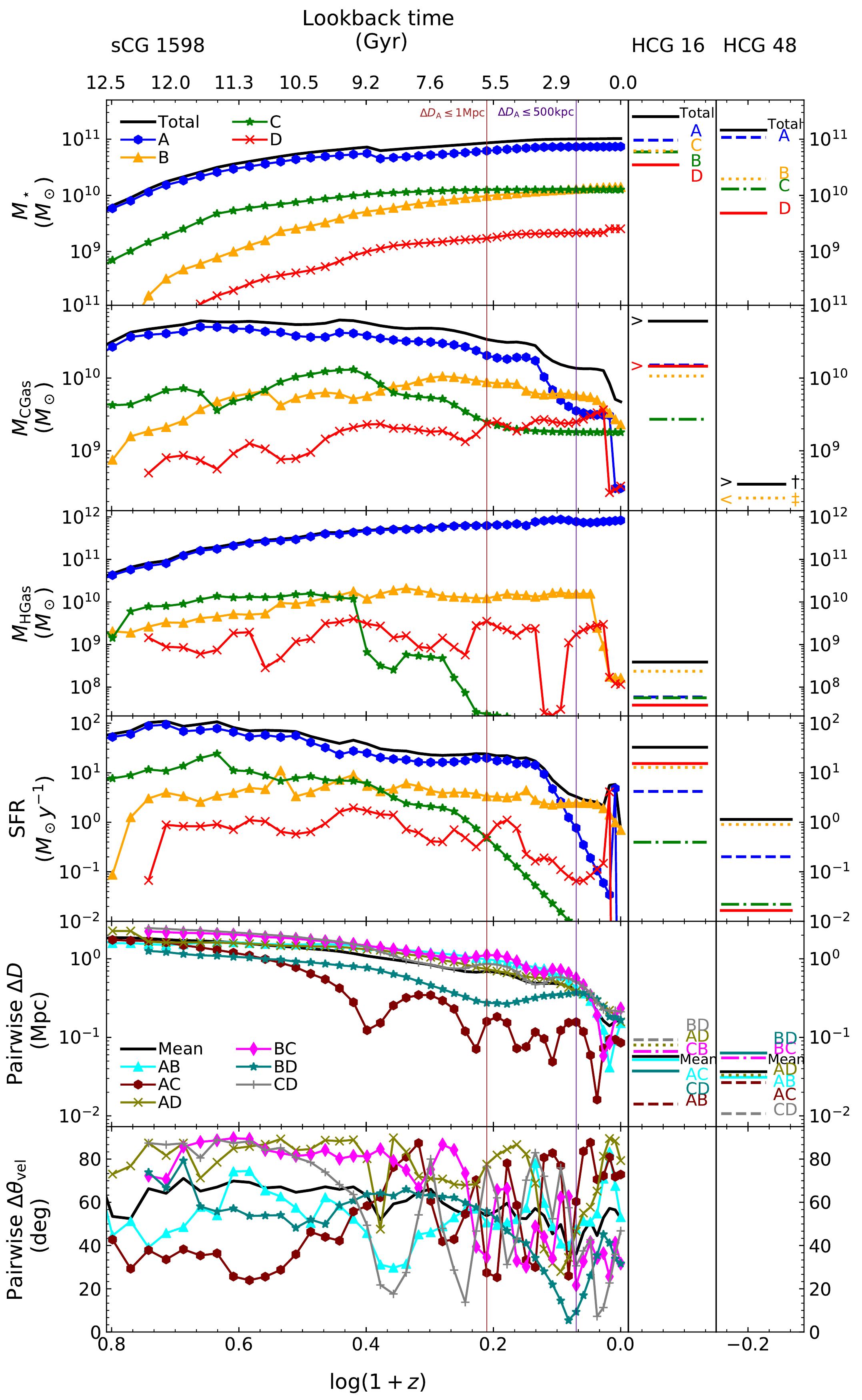}}
    \end{minipage}
    \vspace{-0.4cm}
    \caption{Top left: Composite image of HCG~16 combining \hst\ WFPC2 ($B_{450}$, $V_{606}$, and $I_{814}$) and ESO Multi-Mode Instrument data. Image credit: NASA, ESA, ESO, and J. Charlton ({\href{http://www.spacetelescope.org/news/heic1514/}{http://www.spacetelescope.org/news/heic1514/}). Second from top left: \swift/\uvot\ color image as in \fr{fig-sCG115_vs_HCG22}, but now showing HCG~16. Second from bottom left: Dark Energy camera six filter ($U_{350}$, $G_{475}$, $R_{635}$, $I_{775}$, $Z_{925}$, and $Y_{1000}$) optical image of HCG~48. Image credit: Dane Kleiner (\href{https://www.noao.edu/news/2014/pr1408-images.php}{https://www.noao.edu/news/2014/pr1408-images.php}). Bottom left: \swift/\uvot\ color image as in \fr{fig-sCG115_vs_HCG22}, but now showing HCG~48. Right: Same as \fr{fig-sCG115_vs_HCG22}, but instead showing sCG~1598 vs HCGs~16 and 48. The color-coding for each sCG~1598 galaxy corresponds directly to that in the 3D merger tree shown in \fr{fig-sCG1598_3D}. The two right-most panels show values for the same quantities as in the left hand panels but for HCG~16 and HCG~48 as labeled (\tr{tab-hcgdata}). In the case of HCG~16, galaxies B and C are not in order of decreasing stellar mass; the line color/style still reflects decreasing stellar mass, so sCG~1598B is now matched with HCG~16C, and sCG~1598C with HCG~16B.}}
  \label{fig-sCG1598_vs_HCG16+48}
%  \begin{minipage}[l]{\linewidth}{
%  \hyptarg{f1598-16-48a}{$^{\rm a}$\footnotesize{\href{http://www.spacetelescope.org/news/heic1514/}{http://www.spacetelescope.org/news/heic1514/}}}\\
%  \hyptarg{f1598-16-48b}{$^{\rm b}$\footnotesize{\href{https://www.noao.edu/news/2014/pr1408-images.php}{https://www.noao.edu/news/2014/pr1408-images.php}}} }
%  \end{minipage}
% 
\end{figure*}

\begin{figure*}[ht]
  \centering
  \includegraphics[scale=.42]{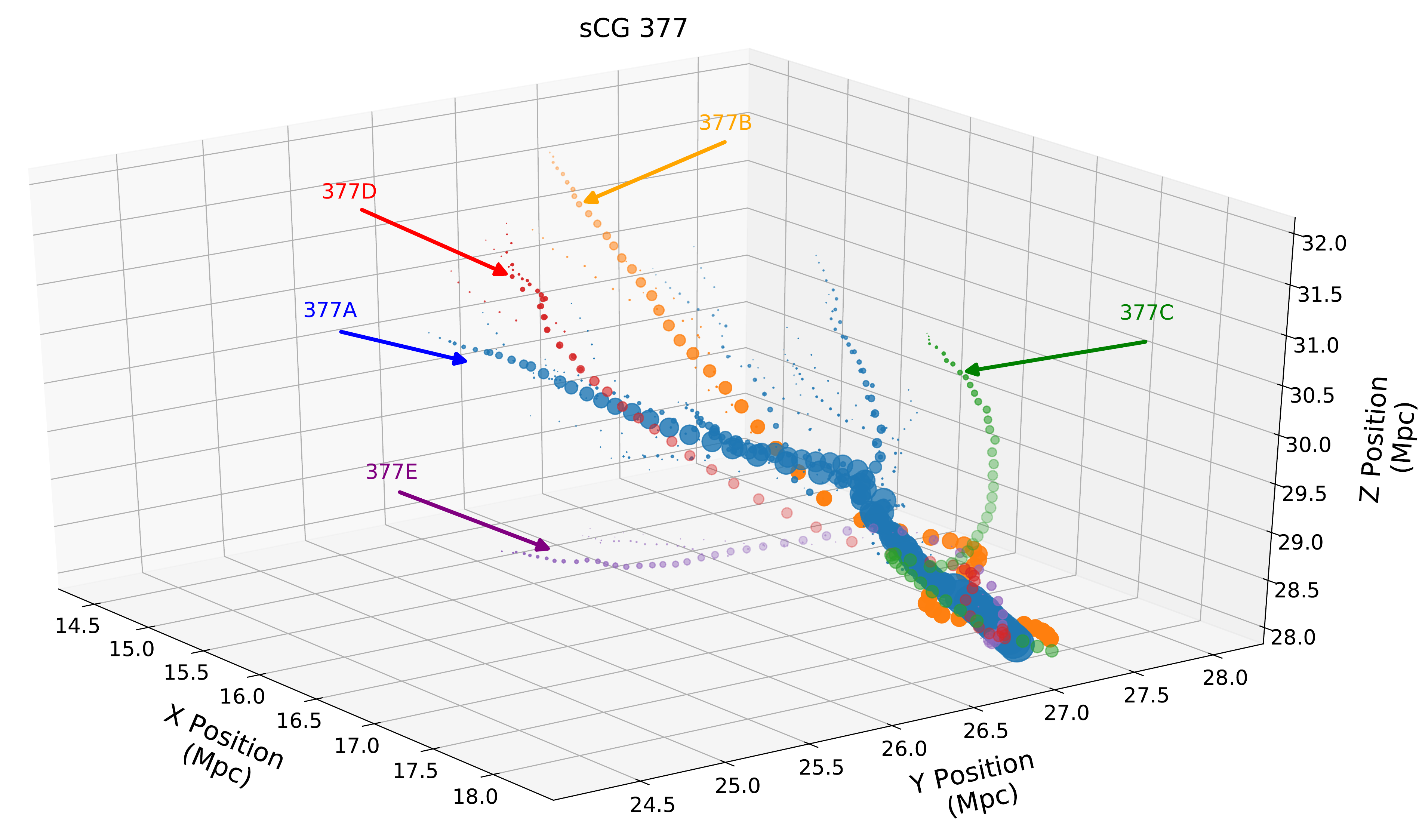}
  \caption{Same as \fr{fig-sCG115_3D}, but now showing sCG~377. The final configuration of the group at $z=0$ is roughly in the lower right of the figure as viewed in perspective. Galaxies are labeled A to E in order of decreasing \mstar\ at $z=0$.} 
  \label{fig-sCG377_3D}
\end{figure*}

\begin{figure*}[ht]
  \centering
    \begin{minipage}{.40\linewidth}
        \centering
        \subfigure{\includegraphics[scale=.562]{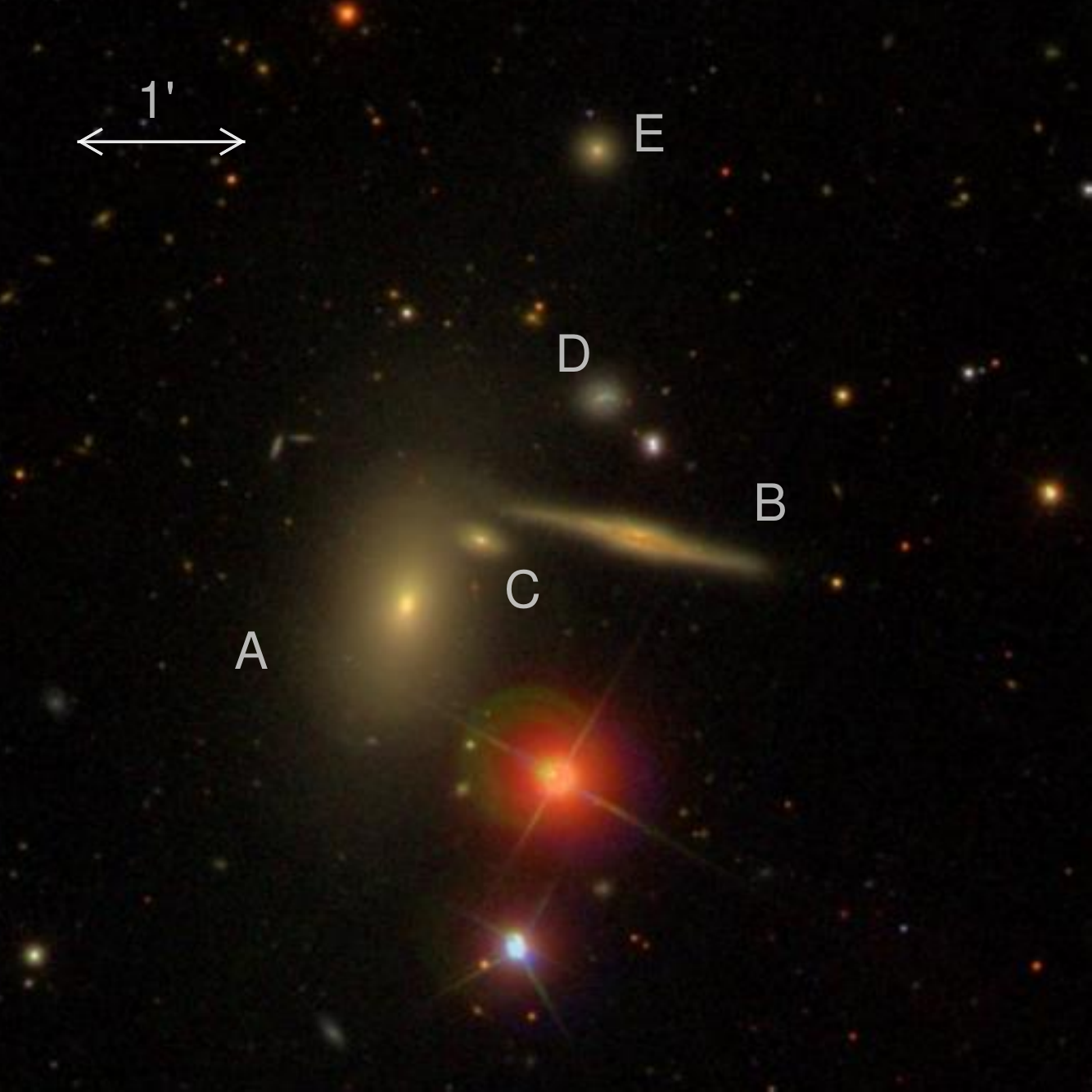}}\\[-.5ex]
        \subfigure{\includegraphics[scale=.252]{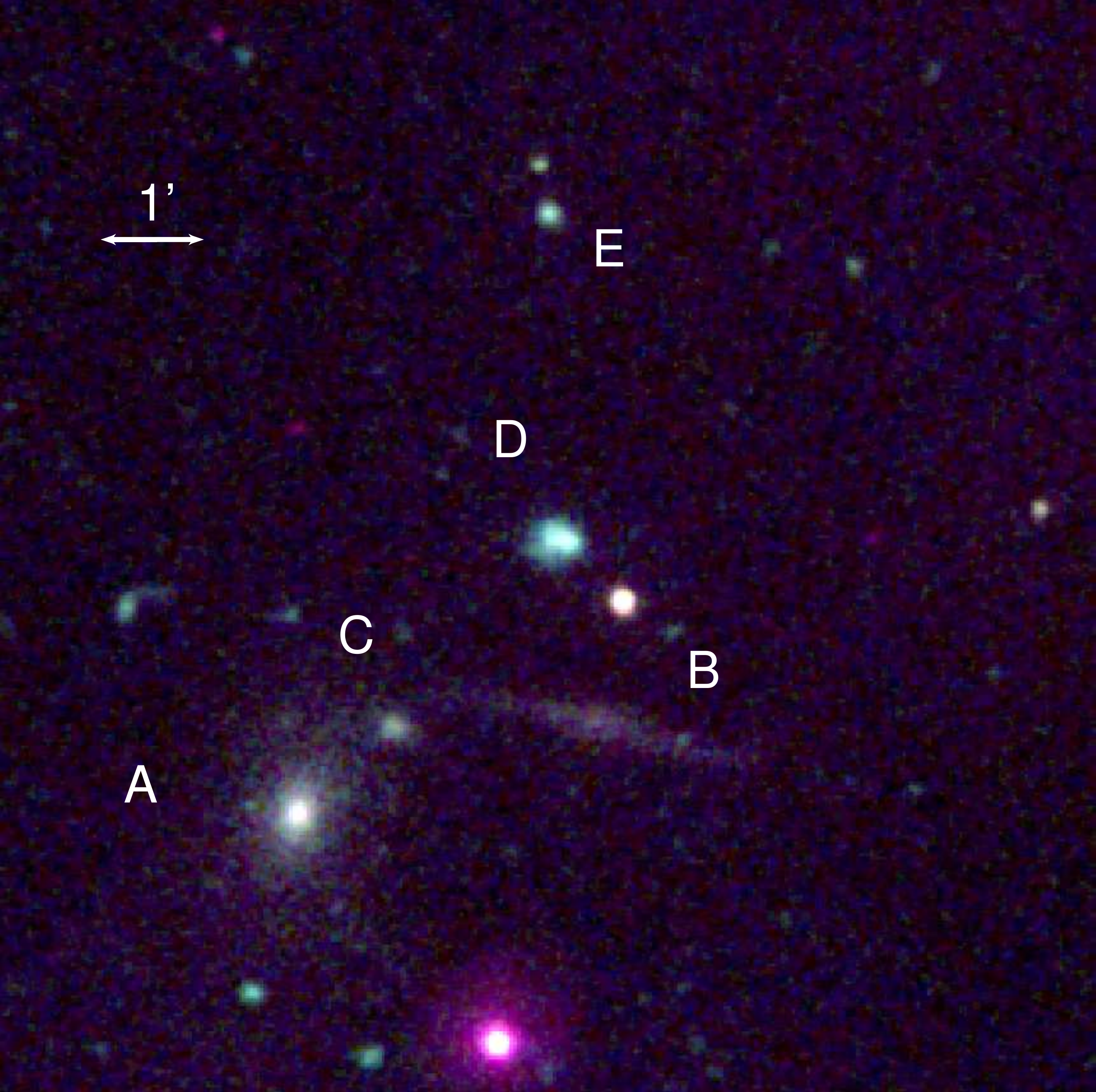}}
        \end{minipage}%
    \begin{minipage}{.595\linewidth}
        \centering
        \subfigure{\includegraphics[scale=.45]{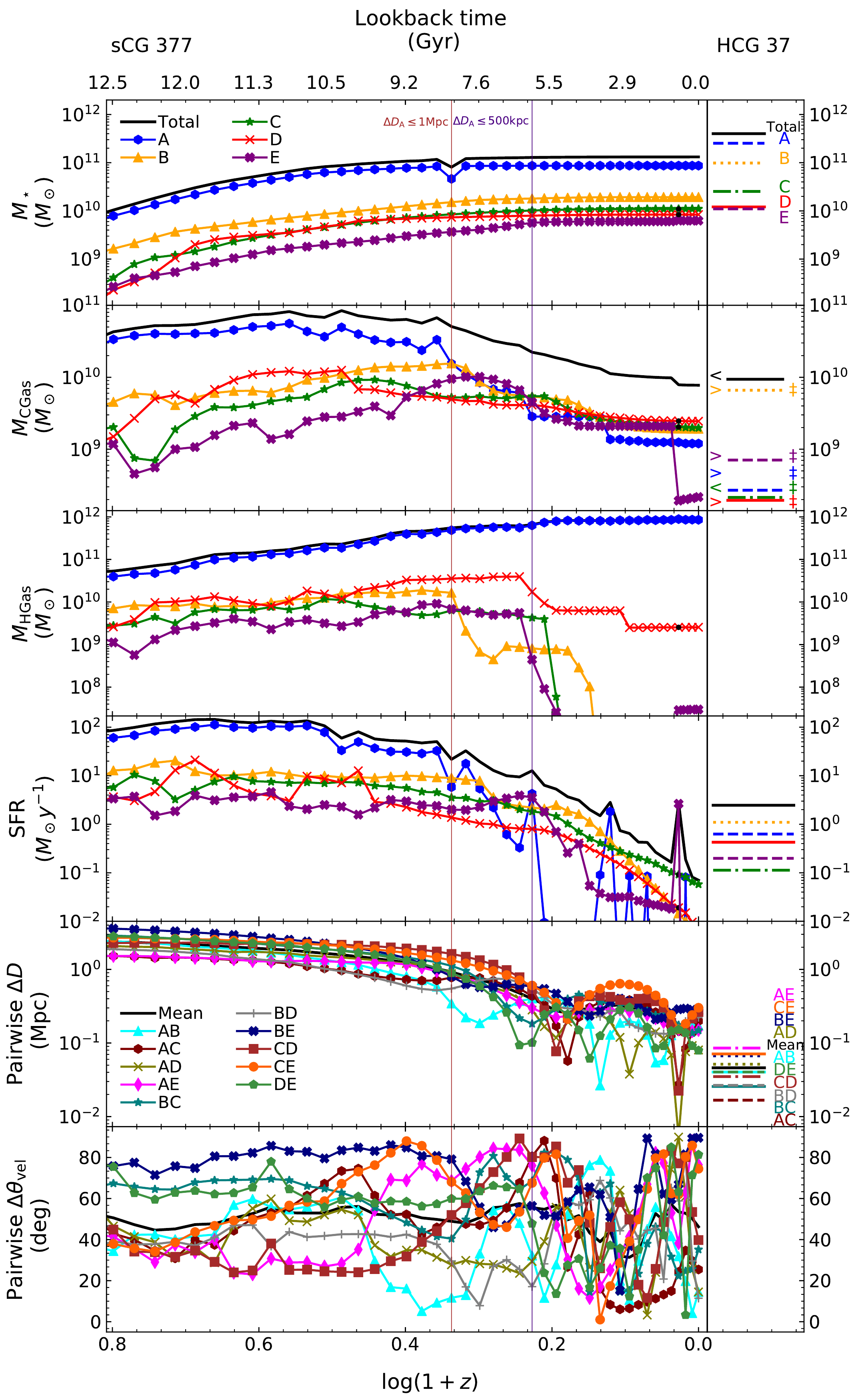}}
    \end{minipage}
   \caption{Top Left: SDSS optical image as in \fr{fig-sCG115_vs_HCG22}, but now showing HCG~37. Bottom Left: \swift/\uvot\ color image as in \fr{fig-sCG115_vs_HCG22}, also instead showing HCG~37. Right: Same as \fr{fig-sCG115_vs_HCG22}, but instead showing sCG~377 vs HCG~37. The color-coding for each sCG~377 galaxy corresponds directly to that in the 3D merger tree shown in \fr{fig-sCG377_3D}. In the bottom two panels, the HCG~37 pairwise separations use the same naming and color- and symbol coding as in the corresponding sCG~377 panels.}
  \label{fig-sCG377_vs_HCG37}
\end{figure*}

For each sCG, we label member galaxies A through E by decreasing stellar mass at $z=0$.
For five SCGs that show the greatest agreement in \mstar\ with observed groups (\scr{sec-compare}),
we present visualizations of the merger trees and evolution of key properties in
Figures~\ref{fig-sCG115_3D}, \ref{fig-sCG115_vs_HCG22}, \ref{fig-sCG1119_3D}, \ref{fig-sCG1119_vs_HCG61}, \ref{fig-sCG1056_3D}, \ref{fig-sCG1056_vs_HCG31}, \ref{fig-sCG1598_3D}, \ref{fig-sCG1598_vs_HCG16+48}, \ref{fig-sCG377_3D} and \ref{fig-sCG377_vs_HCG37}. In the merger tree visualizations, each galaxy is indicated by a different filled colored circle throughout its evolutionary history that also includes mergers; merger constituents retain the $z=0$ galaxy color. Circle sizes scale with relative galaxy stellar mass.

Apart from the properties gleaned from the \mmsg\ catalog for individual galaxies, we also calculated two additional quantities. For each pair of sCG galaxies in each redshift snapshot, we calculated the pairwise Euclidean distance, \dd, between the galaxy positions in 3D space. We separately label \dda\ the distances between the galaxy with the most stellar mass, A, and other sCG member galaxies; for each sCG we identified the redshift snapshots where each galaxy (other than A) comes within \dda~$\leqslant \ 1 \ \rm{Mpc}$ and $\leqslant \ 500 \ \rm{kpc}$ of the most massive group galaxy. The significance of this is discussed below. We also calculated angular separations of velocity vectors for all sCG member galaxy pairs across all snapshots. We defined these vectors for each galaxy by means of start- and end-position vectors for each look-back time snapshot interval. Over cosmic time these provide a quantitative characterization of the directional evolution of each galaxy's path in 3D space. For each galaxy pair we thus obtained the evolution of the angle \dtheta\ between the velocity vectors of the two galaxies by means of the cross product of the vectors. 
%DM: The velocity vector cross products show between every pairwise galaxy set not just between each galaxy and galaxy A. 
Due to merging, in $z>0$ snapshots a single $z=0$ galaxy may correspond to several precursors. In such cases \dd\ and \dtheta\ were calculated between the most massive galaxy precursors of the $z=0$ galaxies.

We present results for these key properties of individual sCG galaxies in \tr{tab-simdata}. Specifically, in Columns (3) to (5) we show values for \mstar, \mcgas, and \mhgas\ both for member galaxies and the group as a whole for each sCG at $z=0$. Columns (7) and (9) give the redshift at which member galaxies come within \dda\ of~1 and 0.5 Mpc, respectively, of the most massive galaxy, A, for the first time.
The significance of these particular values is that, once galaxies come this close, they remain within 1 (0.5) Mpc of A to $z=0$. Thus, these redshifts can be considered as fiducial points in cosmic time when the parent group is ``born," while these separations from galaxy A are in some sense two possible measures of a maximal ``size" for a galaxy concentration to be considered an sCG. Note that this is an empirical result; we also explored smaller separations, but the galaxies did not in general remain within these smaller separations until $z=0$. For example, among the 23 galaxies in all sCGs that are not the most massive ones in their group, only 7 come within 100 kpc of their group's galaxy A and stay within this distance until $z=0$.
Using this redshift information, we calculated and show sCG lifetimes, which are simply the cosmic time that has elapsed between each of these redshift values and $z=0$ (Columns (8) and (10) of \tr{tab-simdata}). Finally, we show total and individual galaxy SFR values at $z=0$ (Column (6)), as well as an average SFR over the lifetime corresponding to separations of 0.5 Mpc (Column (11)).

\vspace{.2cm}
As mentioned, for the five sCGs that show the best agreement with observed HCGs (\scr{sec-compare}), we plot the evolution of these properties in Figures~\ref{fig-sCG115_vs_HCG22}, \ref{fig-sCG1119_vs_HCG61}, \ref{fig-sCG1056_vs_HCG31}, \ref{fig-sCG1598_vs_HCG16+48}, and \ref{fig-sCG377_vs_HCG37}. These figures have a left part, showing archival images of the best-agreement HCGs, and a right part,  to which we will refer hereafter as ``evolutionary figures/plots".  These evolutionary plots consist of a series of left and right panels. The left panels plot the evolution of properties for individual sCG galaxies using different symbols and colors for each. From top to bottom the left panels show the evolution of \mstar, \mcgas, \mhgas, SFR, \dd, and \dtheta. Except for \dd\ and \dtheta, we also show the evolution of the total sCG values with black lines. For \dd\ and \dtheta, the black lines show the evolution of their respective mean values instead. The brown and purple vertical lines indicate redshifts and look-back times for \dda~$\le 1$ and 0.5~Mpc as discussed above. Finally, the top horizontal axis represents look-back time, and the bottom one redshift. The right panels in the evolutionary plots show the same quantities for $z=0$ observed HCGs, where available; they are discussed in \scr{sec-compare}. 

\vspace{.2cm}
In a handful of cases some snapshots had missing data in \mmsg. In
such cases we linearly interpolated between the values in the two
adjacent snapshots. The cause of the missing data is not entirely
clear, but this seems to occur when galaxies come very close to one
another, sometimes resulting in the least massive one not being
detected in a snapshot.  We only performed this interpolation where
one of the main galaxies detected at $z=0$ was missing data at earlier
times and only up to its last merging event. Our goal of tracing $z=0$
galaxies all the way back to the beginning of cosmic time inevitably
means that, due to merging events, one or more of these will have
split up into more than one galaxy at earlier times. In turn, one of
the lower-mass precursor galaxies may be missing data; we do not
interpolate such cases. The effect of this is minimal on the traced
quantities, but it does explain why in a few cases some quantities,
usually for the most massive galaxy A, show abrupt dips, e.g. in
sCG~1056A at around 7.0 and 5.5 Gyr ago in \mstar\ and \mcgas\ (see
\fr{fig-sCG1056_vs_HCG31}).

\renewcommand{\baselinestretch}{0.6}\selectfont

\renewcommand{\baselinestretch}{1}\selectfont

\renewcommand{\baselinestretch}{0.6}\selectfont

\renewcommand{\baselinestretch}{1}\selectfont
\renewcommand{\baselinestretch}{0.6}\selectfont

\renewcommand{\baselinestretch}{1}\selectfont

\begin{deluxetable*}{cccMccMMMMM}[ht]
\tablewidth{0pt}
%use the following array stretch command to scale the graph vertically to fit
%use the later \hspace commands to stretch the horizontal white space for appropriate fit
\renewcommand{\arraystretch}{.80}
\tablecaption{Observational HCG Sample\label{tab-hcgdata}}
\tablehead{
\multicolumn{2}{c}{\hspace{.5cm}Group ID}\hspace{.8cm}
& \multicolumn{2}{c}{\hspace{.9cm}Coordinates (J2000)}\hspace{.5cm}
& \colhead{\hspace{.2cm}$D_{\rm L}$\tm{a}}\hspace{.2cm}
& \colhead{\hspace{.4cm}$M_*$\tm{b}}\hspace{.3cm}
& \colhead{\hspace{.3cm}$M_{{\rm H}{\textsc i}}$\tm{c}}\hspace{.3cm}
& \colhead{\hspace{.3cm}$M_{\rm H_2}$\tm{d}}\hspace{.3cm}
& \colhead{\hspace{.3cm}\mcgas\tm{e}}\hspace{.3cm}
& \colhead{\hspace{.3cm}$M_{\rm H Gas}$\tm{f}}\hspace{.4cm}
& \colhead{\hspace{.4cm}SFR\tm{b}}\hspace{.4cm}
\\
\multicolumn{11}{c}{\vspace{-0.7cm}}
\\
\multicolumn{2}{c}{\rule{2.25cm}{.03cm}}
& \multicolumn{2}{c}{\rule{4.0cm}{.03cm}}
& \colhead{}
& \multicolumn{5}{c}{\rule{7.5cm}{.03cm}}
& \colhead{}
\\
\multicolumn{11}{c}{\vspace{-0.6cm}}
\\
\multicolumn{2}{c}{}
& \colhead{$\alpha$}
& \colhead{$\delta$}
& \colhead{(Mpc)}
& \multicolumn{5}{c}{$\log(M_\odot)$}
& \colhead{$\log(M_\odot $ ${\rm yr}^{-1})$}
\\
\colhead{(1)}
& \colhead{(2)}
& \colhead{(3)}
& \colhead{(4)}
& \colhead{(5)}
& \colhead{(6)}
& \colhead{(7)}
& \colhead{(8)}
& \colhead{(9)}
& \colhead{(10)}
& \colhead{(11)}
}
\startdata
HCG~16 & Group     & 02\rah09\ram31.3\ras & -10^\circ09\arcmin31\arcsec & 51.3 & 11.41 & >10.44       & 10.52       & >10.78          & 8.59\substack{+8.35 \\ -8.30} &  1.52 \\ 
       & A         & 02\rah09\ram24.6\ras & -10^\circ08\arcmin09\arcsec &      & 10.98 &   9.09       & 10.14       &  10.18          & 7.77\substack{+7.52 \\ -7.52} &  0.62 \\
       & B         & 02\rah09\ram20.8\ras & -10^\circ07\arcmin59\arcsec &      & 10.77 &   8.92       &  9.27       &   9.43          & 7.75\substack{+7.75 \\ -7.22} & -0.40 \\
       & C         & 02\rah09\ram38.5\ras & -10^\circ08\arcmin48\arcsec &      & 10.79 &   9.50       &  9.87       &  10.02          & 8.37\substack{+8.09 \\ -8.13} &  1.11 \\
       & D         & 02\rah09\ram42.9\ras & -10^\circ11\arcmin03\arcsec &      & 10.54 &  >9.67       &  9.99       & >10.16          & 7.58\substack{+7.14 \\ -7.14} &  1.19 \\
HCG~22 & Group     & 03\rah03\ram31.3\ras & -15^\circ40\arcmin32\arcsec & 34.8 & 10.99 &   9.15       & \ldots      &  >9.15 \dagger  & \ldots                        & -0.08 \\
       & A         & 03\rah03\ram38.4\ras & -15^\circ36\arcmin48\arcsec &      & 10.92 &  \ldots      &  <8.82      &  <8.82 \ddagger & \ldots                        & -0.56 \\
       & B         & 03\rah03\ram26.1\ras & -15^\circ39\arcmin43\arcsec &      &  9.84 &  \ldots      & \ldots      &  \ldots         & \ldots                        & -1.30 \\
       & C         & 03\rah03\ram24.5\ras & -15^\circ37\arcmin24\arcsec &      &  9.86 &  \ldots      &  <8.70      &  <8.70 \ddagger & \ldots                        & -0.30 \\
HCG~31 & Group     & 05\rah01\ram38.3\ras & -04^\circ15\arcmin25\arcsec & 55.7 & 10.49 &  10.37       & <9.28       & <10.40          & \ldots                        &  1.07 \\ 
       & ACE\tm{g} & 05\rah01\ram38.7\ras & -04^\circ15\arcmin34\arcsec &      & 10.32 &   9.67\tm{h} & <8.84\tm{h} &  <9.73          & \ldots                        &  0.96 \\
       & B         & 05\rah01\ram36.2\ras & -04^\circ15\arcmin43\arcsec &      &  9.52 &   9.30       &  <8.80      &  <9.42          & \ldots                        & -0.01 \\
       & G         & 05\rah01\ram44.0\ras & -04^\circ17\arcmin20\arcsec &      &  9.81 &   9.30       &  <8.75      &  <9.41          & \ldots                        &  0.22 \\
HCG~37 & Group     & 09\rah13\ram35.6\ras & +30^\circ00\arcmin51\arcsec & 96.8 & 11.17 &   9.21       & <9.89\tm{i} &  <9.97          & \ldots                        &  0.26 \\
       & A         & 09\rah13\ram39.4\ras & +29^\circ59\arcmin35\arcsec &      & 11.40 &  \ldots      &  8.43\tm{i} &  >8.43 \ddagger & \ldots                        & -0.20 \\
       & B         & 09\rah13\ram33.1\ras & +30^\circ00\arcmin01\arcsec &      & 10.99 &  \ldots      &  9.82\tm{i} &  >9.82 \ddagger & \ldots                        &  0.04 \\
       & C         & 09\rah13\ram37.3\ras & +29^\circ59\arcmin58\arcsec &      & 10.40 &  \ldots      & <8.33\tm{i} &  <8.33 \ddagger & \ldots                        & -0.95 \\
       & D         & 09\rah13\ram33.8\ras & +30^\circ00\arcmin57\arcsec &      & 10.08 &  \ldots      &  8.29\tm{i} &  >8.29 \ddagger & \ldots                        & -0.37 \\
       & E         & 09\rah13\ram34.0\ras & +30^\circ02\arcmin23\arcsec &      & 10.04 &  \ldots      &  8.85\tm{i} &  >8.85 \ddagger & \ldots                        & -0.70 \\ 
HCG~48 & Group     & 10\rah37\ram45.6\ras & -27^\circ04\arcmin50\arcsec & 43.7 & 11.16 &   8.54       & \ldots      &  >8.54 \dagger  & \ldots                        &  0.06 \\
       & A         & 10\rah37\ram47.4\ras & -27^\circ04\arcmin54\arcsec &      & 11.03 &  \ldots      & \ldots      &  \ldots         & \ldots                        & -0.70 \\ 
       & B         & 10\rah37\ram49.5\ras & -27^\circ07\arcmin18\arcsec &      & 10.29 &  \ldots      &  <8.35      &  <8.35 \ddagger & \ldots                        & -0.04 \\
       & C         & 10\rah37\ram40.5\ras & -27^\circ03\arcmin29\arcsec &      & 10.11 &  \ldots      & \ldots      &  \ldots         & \ldots                        & -1.66 \\ 
       & D         & 10\rah37\ram41.4\ras & -27^\circ02\arcmin40\arcsec &      &  9.68 &  \ldots      & \ldots      &  \ldots         & \ldots                        & -1.78 \\
HCG~61 & Group     & 12\rah12\ram24.9\ras & +29^\circ11\arcmin21\arcsec & 58.0 & 11.30 &  9.98        & \ldots      &  >9.98 \dagger  & \ldots                        &  0.69 \\ 
       & A         & 12\rah12\ram18.8\ras & +29^\circ10\arcmin46\arcsec &      & 11.03 &  \ldots      &  8.42       &  >8.42 \ddagger & \ldots                        & -0.22 \\
       & C         & 12\rah12\ram31.0\ras & +29^\circ10\arcmin06\arcsec &      & 10.82 &  \ldots      &  9.69       &  >9.69 \ddagger & \ldots                        &  0.61 \\
       & D         & 12\rah12\ram26.9\ras & +29^\circ08\arcmin57\arcsec &      & 10.43 &  \ldots      & \ldots      &  \ldots         & \ldots                        & -0.67 \\
\enddata
\vspace{.1cm}
\hspace{-.5cm}
\begin{minipage}[0.01\textheight]{1.05\textwidth}
  {\footnotesize
{\bf Notes.} Member galaxy properties for HCGs used in case studies.
Column (1): CG name. Column (2): member galaxy ID. Column (3): R.A. Column (4): decl. Column (5): luminosity distance from \href{https://ned.ipac.caltech.edu/}{NED}$^{\rm a}$. Column (6): stellar mass. Column (7): \hone\ mass. Column (8): \htwom\ mass. Column (9): hot gas mass. Column (10): cold gas mass. Column (11): star formation rate. Group total values are sums of individual galaxy values, except where noted.

      \vspace{0cm}
      \txs{a} \href{https://ned.ipac.caltech.edu/}{https://ned.ipac.caltech.edu/}
      \vspace{0cm}

      \txs{b} \cite{lenkic2016}. 
      \vspace{0cm}

      \txs{c} \citet{verdes-montenegro2001}. Much of the total group \hone\ mass is not confined to individual galaxies and is thus not the sum of individual galaxy values.
      \vspace{0cm}

      \txs{d} \citet{lisenfeld2017}, except where noted.
      \vspace{0cm}

      \txs{e} \mcgas\ calculated as the sum of \hone\ and \htwom\ gas masses, where available. The $\dagger$ indicates cases where only \mhone\ data were available, while the $\ddagger$ indicates the same for \mhtwom. In cases where the only available data are an upper limit for \mhtwom, this is indicated by $<$ and a $\ddagger$ in the \mcgas\ Column (9).
      \vspace{0cm}

      \txs{f} \citet{osullivan2014a}.
      \vspace{0cm}

      \txs{g} Galaxies A, C, and E treated as a single entity by references. 

      \vspace{0cm}

      \txs{h} Data for AC only.
      \vspace{-.1cm}

      \txs{i} \mhtwom\ values for HCG~37 from ~\citet{martinez-badenes2012}.
  }
\end{minipage}
\vspace{-1cm}
\end{deluxetable*}

\renewcommand{\baselinestretch}{0.6}\selectfont
\renewcommand{\baselinestretch}{1}\selectfont
\renewcommand{\baselinestretch}{.6}\selectfont

\renewcommand{\baselinestretch}{1}\selectfont
\renewcommand{\baselinestretch}{.6}\selectfont
\renewcommand{\baselinestretch}{1}\selectfont
\renewcommand{\baselinestretch}{.6}\selectfont

\renewcommand{\baselinestretch}{1}\selectfont
\renewcommand{\baselinestretch}{.6}\selectfont
\begin{figure*}
%\gridline{\fig{fig-histmstar-f.pdf}{0.4\textwidth}{(a)}
 %         \fig{fig-histsfr-f.pdf}{0.4\textwidth}{(b)}
  %        \fig{fig-dsep-f.pdf}{0.4\textwidth}{(b)}
%}
%\includegraphics[scale=0.6]{fig-hists-f.pdf}\vspace{-1cm}
\begin{minipage}[c]{0.3\linewidth}
\hspace{-0.8cm}
\includegraphics[scale=0.15]{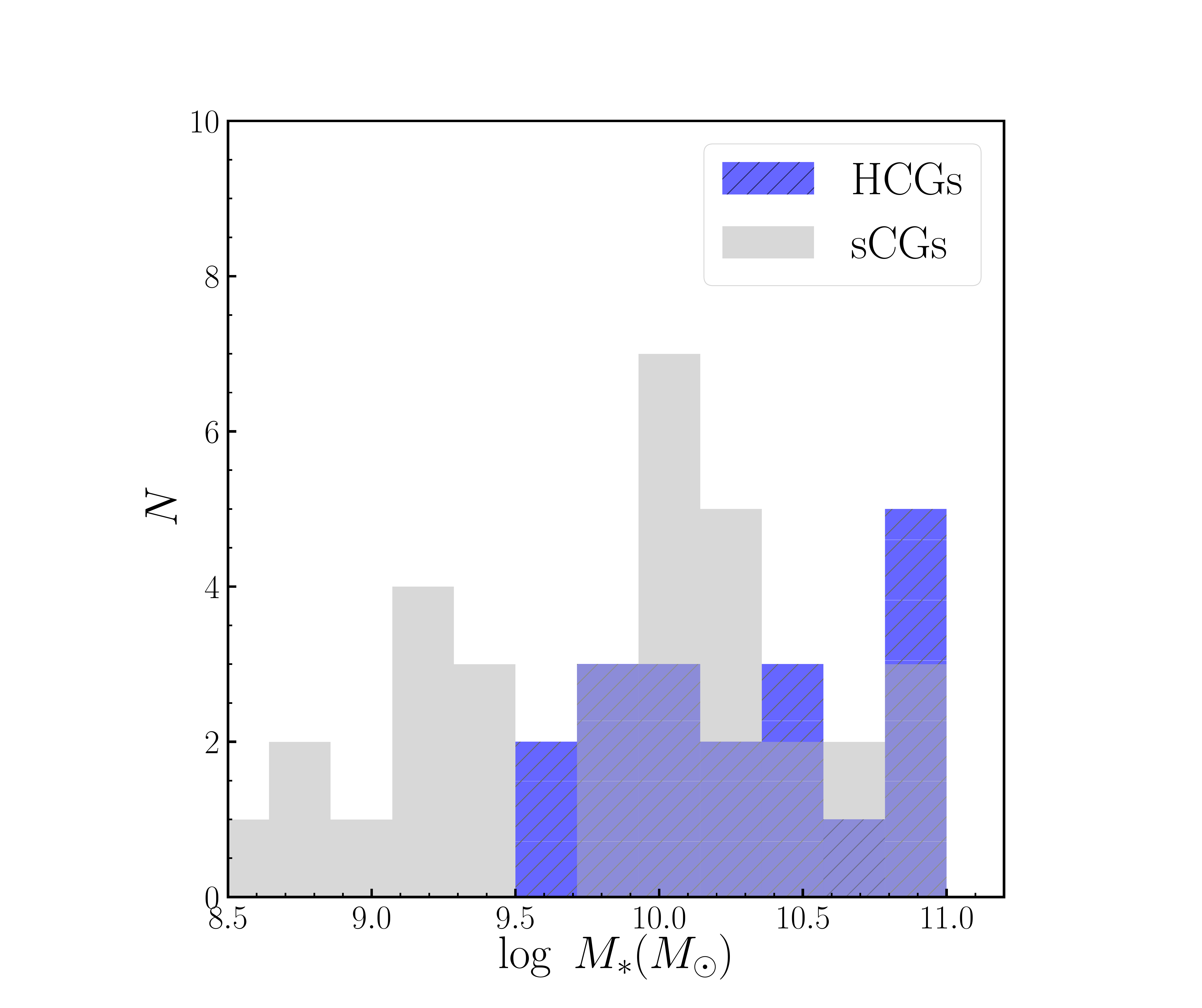}
\end{minipage}
\begin{minipage}[c]{0.3\linewidth}
\includegraphics[scale=0.15]{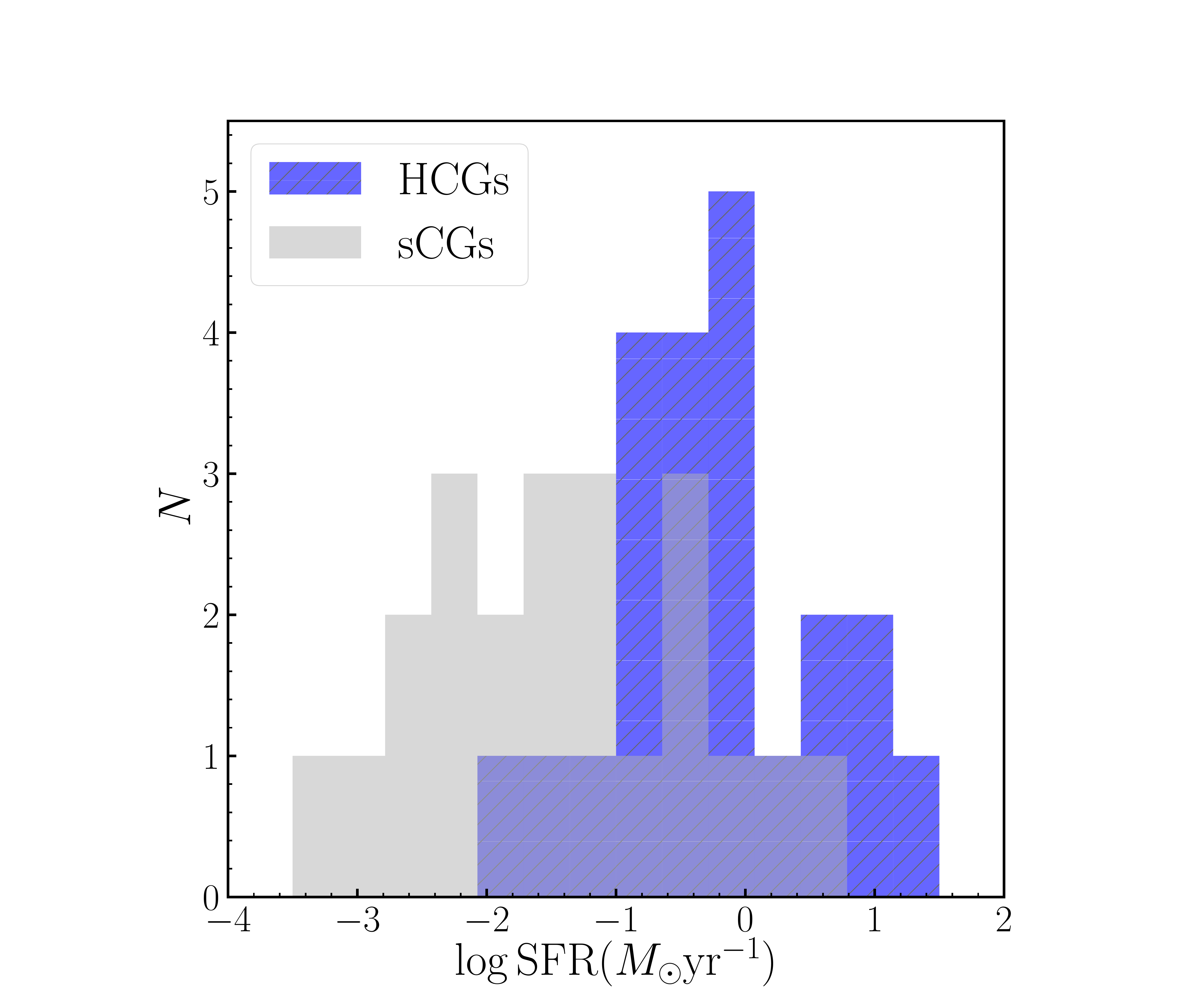}
\end{minipage}
\begin{minipage}[c]{0.3\linewidth}
\hspace{+0.6cm}
\includegraphics[scale=0.15]{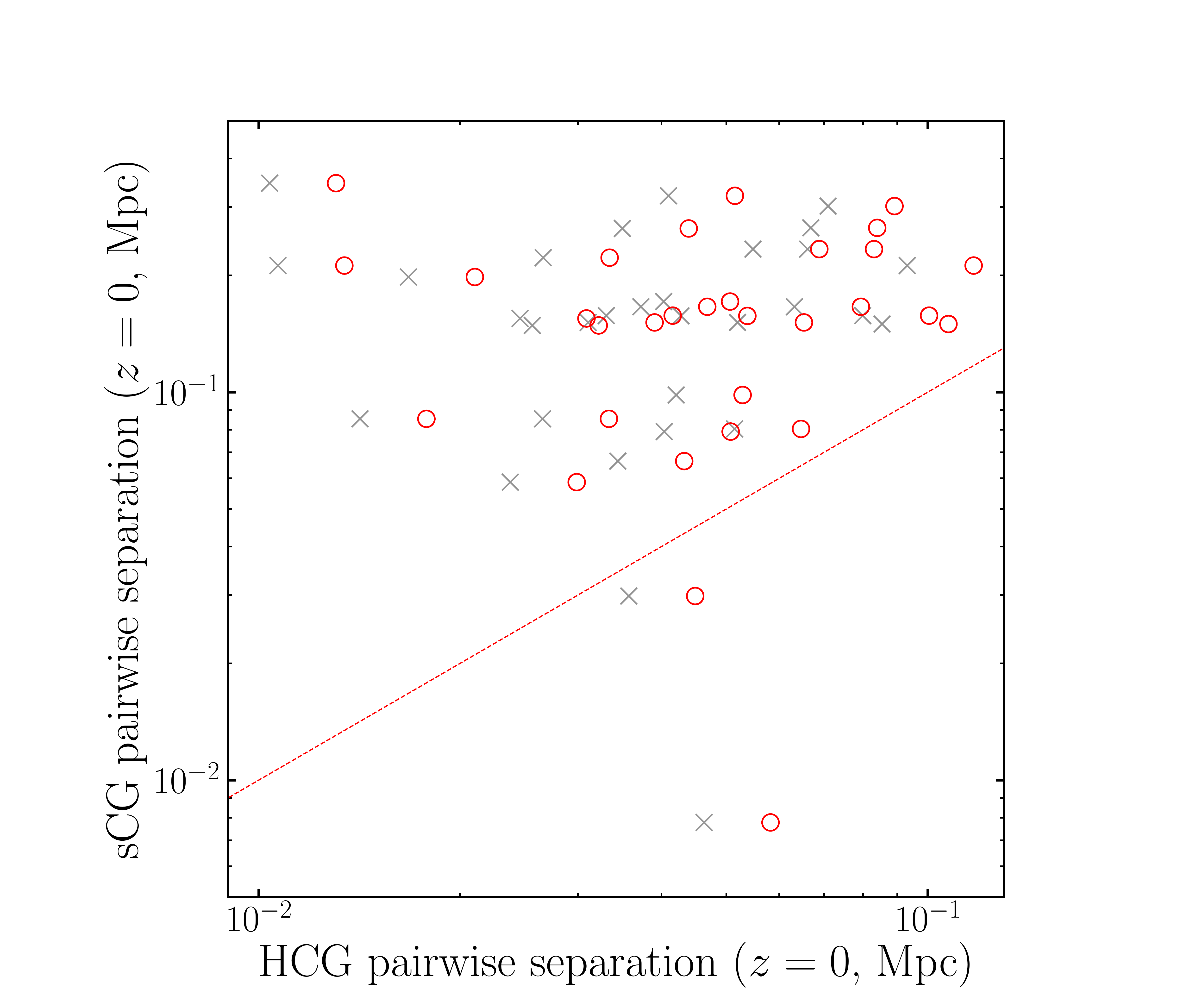}
\end{minipage}
\caption{Distributions of \mstar\ (left) and SFR (middle) values for HCG (blue hatched) and sCG (gray) galaxies. The right panel shows sCG pairwise separations at $z=0$ against HCG pairwise separations for corresponding, \mstar-ranked galaxies. The gray crosses are original projected separations, while red circles mark these separations multiplied by an empirically determined factor of $2\pi/5$ to ``correct'' them to reasonable deprojected separations. The red dashed line indicates  equality.\label{fig-hists}}
%Below will make custom references right after caption to circumvent duplication bug.
%\begin{minipage}{0.3\linewidth}
%{$^{123}$This is a test.}
%\end{minipage}\\
%\begin{minipage}{0.3\linewidth}
%{$^{4}$This is another test.}
%\end{minipage}
\end{figure*}

\renewcommand{\baselinestretch}{1}\selectfont
\renewcommand{\baselinestretch}{.6}\selectfont
\begin{figure}
  %\centering
  %\hspace{-1.cm}
  \includegraphics[trim={150 0 0 105},clip,scale=.23]{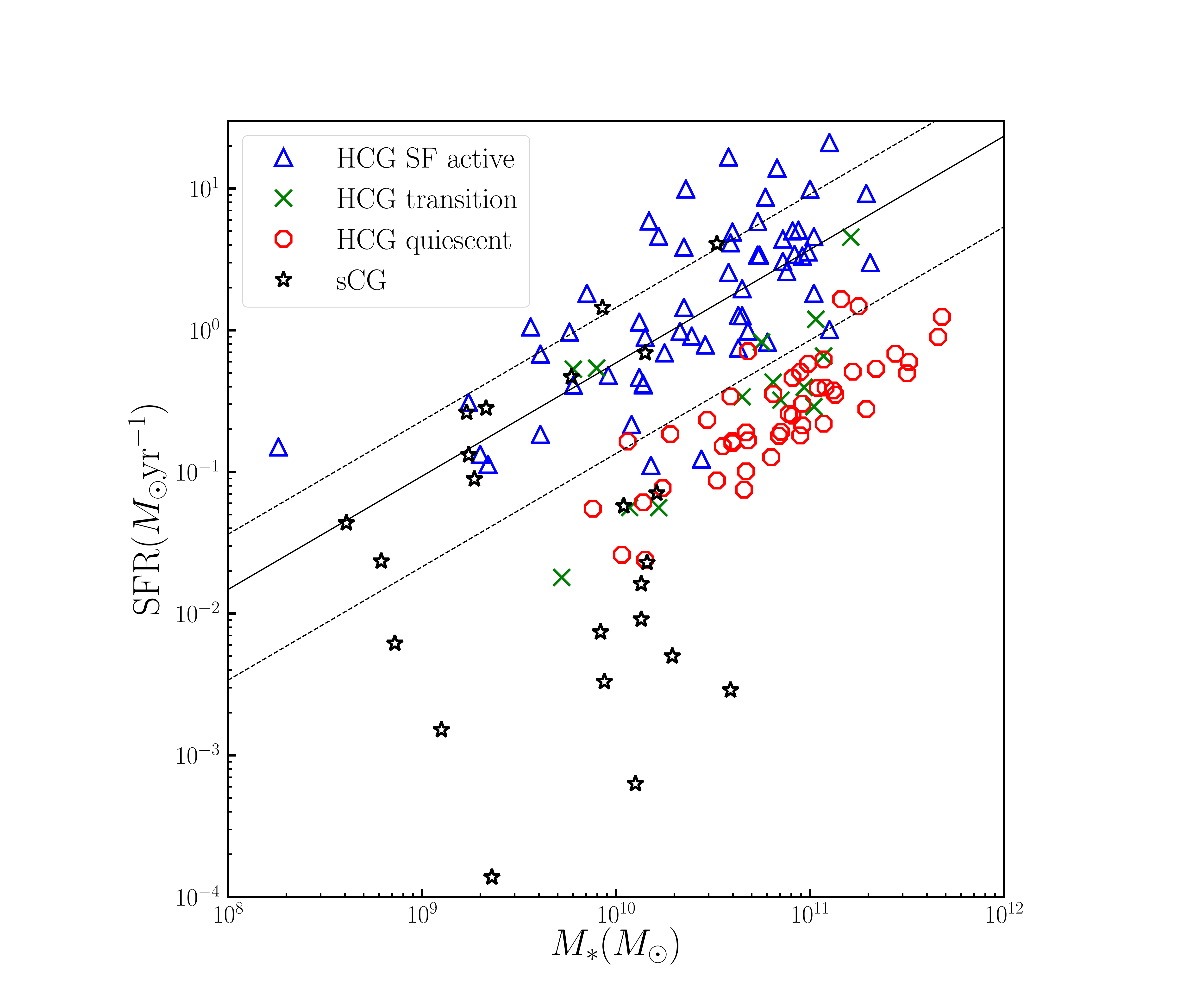}
  \caption{SFR vs.~\mstar\ plot for CG galaxies in this paper, adapted from \citet{lenkic2016}. We used classification information from \citet{zucker2016} to color-code HCGs according to their level of star-formation activity: blue triangles are actively star-forming, green crosses are transitioning, and red circles indicate quiescent galaxies. Black stars represent sCG galaxies from our study. The black line is the ``star-forming main sequence'' fit of \citet{chang2015}. The dotted lines indicate $\pm1\sigma$ limits.}
  \label{fig-sfms}
\end{figure}

\renewcommand{\baselinestretch}{1}\selectfont

\subsection{Results}
We first describe results for the three different types of galaxy masses (\mstar, \mcgas, and \mhgas, all at $z=0$) 
shown in \tr{tab-simdata}, which show some clear and interesting patterns.

\vspace{.2cm}
Among three-member systems, most show similar 
stellar-mass ranges at $z=0$: the stellar mass for their most massive galaxy is in the range $10^{10} \lesssim M_{\ast}/M_{\odot} \lesssim 10^{11}$. In comparison, the stellar mass of the second most massive galaxy is smaller by less than an order of magnitude. That said, member galaxies in sCG~518, sCG~1773, and sCG~2143 have \mstar\ values up to $\sim$1 order of magnitude smaller.

Regardless of the number of members, in all systems the cold gas mass shows a variable range, but its total value never exceeds $10^{10}$~\msun, which is well below the total stellar mass. The only exception is sCG~518, where \mcgas~$=1.8$~\mstar. 
 In terms of \mcgas\ in {\it individual} galaxies and in triplets, in half of triplet sCGs the two galaxies that are least massive in \mstar\ (i.e., B and C) are more massive in \mcgas\ compared to galaxy A. The exceptions are sCG~1441~C, sCG~1757~B, and both B and C in sCG~518, sCG~1773, and sCG~2143. In contrast, with the exception of sCG~518, the total hot gas mass is consistently larger by up to about an order of magnitude than the total stellar mass. This is driven by the hot gas mass associated with the galaxy that has the highest \mstar\ (i.e., A). In most three-member sCGs galaxy A is also the galaxy with the highest \mhgas, the exceptions being sCG~518 and sCG~1441. 

In the single four-member sCG~1598, \mstar\ ranges from $10^{9.4}$ to $10^{11}$~\msun, and galaxy A has the least cold gas and the most hot gas mass, completely dominating the total hot gas mass. The five-member sCG~377 is similar: \mstar\ ranges from $10^{9.8}$ to $10^{11.12}$~\msun, and galaxy A is the second least massive galaxy in \mcgas\ while it provides the dominant contribution by more than three orders of magnitude to the group's total hot gas mass. 

Regarding the SFR values at $z=0$, overall these show relatively low values even if we exclude galaxies to which the Guo2010a semianalytical model assigns SFR values of zero.
Comparing Columns (6) and (11) in \tr{tab-simdata}, we see that, overall, SFRs for sCG galaxies were higher in the past, as expected. 
The only exception is sCG~1757, for which the mean SFR for $z\lesssim 0.28$ is almost an order of magnitude lower than at $z=0$. This is due to a spike in SFR at $z\sim0$ for most massive galaxy A that dominates the SFR at $z=0$, and is due to a merger. Otherwise, a trend with SFR increasing with redshift also holds for this sCG. For the five sCGs that are most similar to HCGs in terms of \mstar, it can be seen from the evolutionary plots that the most massive galaxies had SFRs about 100 times higher close to the beginning of cosmic time. Nevertheless, the general trend of decreasing SFRs is often interrupted by abrupt increases lasting several hundred Myr.

As can be seen in the evolutionary plots, there is considerable variety in the evolution of galaxy separations in the identified sCG sample. As mentioned already, once individual galaxies come within 1 or 0.5 Mpc of the most massive galaxy A, they stay within this distance to A until $z=0$. Considering separations between {\it all} pairs of galaxies, in sCG~115, sCG~377, sCG~1441, and sCG~1598, although at some point galaxies come as close as $\sim$100 kpc or less to each other, they then increase their separations up to $\sim$400 kpc. On the other hand, galaxies in other sCGs do not show this behavior and continuously come closer to each other. A comparison of the lower two left hand panels in the evolutionary plots shows that the degree of increase in \dd\ after close approaches strongly correlates with the relative direction from which the group member galaxies are coming together. Thus, groups where the galaxies are coming together on fairly parallel paths (low \dtheta\ values) do not experience large spikes in pairwise separation after close approaches, while groups whose member galaxies are coming from very different directions (high \dtheta) tend to experience more pronounced and abrupt changes in galaxy separations. 
%
%
    %
    %
    %
    %
    %
    %
    %
    %
    %
    %
%\end{itemize}

We highlight some specific cases for \dd\ and \dtheta\ evolution of particular interest. In sCG 377, which is the only five-member group (Figs~\ref{fig-sCG377_3D} and \ref{fig-sCG377_vs_HCG37}), galaxy B has a closest approach of galaxy A at 57 kpc at $z=0.51$. It then recedes as far as 394 kpc from A at $z=0.28$. The closest the galaxies get to each other after this is at $z=0.04$ when they are 83 kpc from each other. 
This is in contrast to the general observation that, overall, sCG galaxies stay within 250 kpc after a first close encounter.
In contrast, sCG 1056 A and sCG 1056 B come within 100 kpc at $z=0.28$ and remain within this separation until $z=0$, with several close approaches between 12 and 30 kpc.
This suggests that these galaxies may be about to merge into a single system. On the other hand, sCG 1056 C remains more than 300 kpc away, although steadily approaching the other two members. In sCG 1119, galaxies A and B appear to be undergoing an ongoing merging process, with separations within 10 kpc in the past 540 Myr. Their first closest approach at 26 kpc is at $z=0.32$, and their maximum separation after this is 44 kpc with several closer approaches until $z=0$.
Similarly, galaxies A and C in sCG 1773 have been within 30 kpc of each other for the past 3.5 Gyr (since $z=0.76$) with multiple close encounters within $\lesssim 10$ kpc. At the same time, sCG 1773 B is separated by at most 34 kpc from A or C during the past 1.1 Gyr, with a couple of close approaches between 3 and 14 kpc.
In fact, this is the only sCG where by $z=0.24$ and all the way to $z=0$ {\it all} galaxies are within 100 kpc of each other with several close approaches.  This system is thus a clear candidate for merging into a single galaxy. Alternatively, this sCG could turn into a so-called ``fossil group''. Such groups have also been identified in the Millennium Simulation \citep{farhang2017}.
Fossil groups (\citealt{ponman1994}; see also \citealt{dariush2010,aguerri2011,farhang2017,kundert2017} and references therein) are characterized by an extended X-ray halo and are dominated by a large early-type galaxy, with a significantly fainter second-ranked galaxy  \citep[at least e mag fainter in $r$ within $R_{200}$, the radius  of the sphere centered on the halo in which the critical density is 200 times the mean density of
the universe;][]{jones2003}. sCG 1773 could become a fossil group if at some point the second most massive (and presumably second-brightest) galaxy B merges with the most massive (and brightest) galaxy A, thus creating a large magnitude gap between the least massive (and least bright) galaxy C and the newly merged galaxy A+B.

\vspace{.2cm}
In terms of angular separation evolution, the mean angular separation for the triplet sCG 115 is $\sim 40$\degr\ until $\sim 6$ Gyr ago. It then fluctuates by up to $\sim 60\%$ until $z=0$. For sCG 1119 the mean \dtheta\ value is $\sim 20$\degr\ in early times, slowly increasing until $\sim 4$ Gyr ago, after which it increases significantly more rapidly until $z=0$. In sCG 1056 the mean angular separation is initially $\sim 40$\degr, decreasing down to $\sim 20$\degr\ by $\sim 11$ Gyr ago. It remains at similarly small values until $z=0$ with a pronounced spike $\sim 2.9$ Gyr ago. For sCG 1598 the mean of $\sim 60$\degr\ is fairly stable until $\sim 2$ Gyr ago, but there are $\sim30\%$ fluctuations for individual galaxies. sCG 377 shows similar behavior with a mean of $\sim 50$\degr. There is thus noticeably different behavior in the cosmic evolution of the velocity vector angles between three-member and four- or five-member sCGs: While triplets show, on average, angles between 20\degr\ and 40\degr, the two sCGs with more members show larger angles. In addition, \dtheta\ fluctuations for individual galaxies are clearly more pronounced for the nontriplets. This suggests that, overall, the paths of triplets are, in comparison, much more undisturbed and closer to being parallel to each other.  We postulate that the relative alignment of the galaxy velocity vectors allows these systems to be relatively long-lived.  Furthermore, these long-lived triplets may be moving along narrow filaments, thus ensuring that they are likely to remain isolated.

\vspace{.2cm}
\section{Comparisons with Observed Compact Groups}\label{sec-compare}
\vspace{.1cm}
To compare properties between our detected sCGs and CGs in the Hickson catalog \citep{hickson1982,hickson1992}, we used SFR and \mstar\ values from \citet{lenkic2016}, \mhone\ from \citet{verdes-montenegro2001}, \mhtwom\ from \citet{lisenfeld2017} and \citet{martinez-badenes2012}, and \mhgas\ values from \citet{osullivan2014a}. We plot sCG and corresponding HCG properties as a function of redshift and cosmic time side by side in the evolutionary Figures \ref{fig-sCG115_vs_HCG22}, \ref{fig-sCG1119_vs_HCG61}, \ref{fig-sCG1056_vs_HCG31}, \ref{fig-sCG1598_vs_HCG16+48}, and \ref{fig-sCG377_vs_HCG37}. Specifically, we plot all key quantities for sCG 115, sCG 1119, sCG 1056, sCG 1598, and sCG 377, together with corresponding values for HCG 22, HCG 61, HCG 31, HCG 16, and HCG 48\footnote{Both HCGs show good \mstar\ agreement with sCG 1598.} and HCG 37, respectively. In addition, in \fr{fig-hists} we show \mstar\ and SFR distributions, and we also plot pairwise velocity separations for HCG and sCG galaxies corresponding in stellar mass. For reference, we also provide a list of observational data and key properties for these six HCGs in \tr{tab-hcgdata}, for both individual galaxies and entire groups. We first present an overview of these comparisons and then discuss individual sCG-HCG pairs.

\vspace{.2cm}
Overall, we found that stellar mass is the property for which the greatest number (six) of well-known observed HCGs show reasonable to good agreement with at least one of the five sCGs plotted in the evolutionary figures. The agreement is within factors of a few for at least one member galaxy. This includes both the four- and five-member sCGs, as well as sCG 1056, which bears some particularly interesting similarities to HCG 31, a well-known and unusual system \citep[e.g.,][]{gallagher2010}. However, although sCG stellar masses cover the full range of HCG values (\fr{fig-hists}),  we note that in 13 out of 17 (76\%)  of the HCG galaxies in the six groups compared, \mstar\ is higher by up to $\sim0.9$ in the log (with a mean of $\sim 0.5$) compared to the corresponding \mstar-ranked sCG galaxy.

\vspace{.2cm}
Further, HCGs have systematically higher SFRs at $z=0$. This is the case not only in terms of total (group) rates but also usually when comparing corresponding \mstar-ranked galaxies. It is also true regardless of whether an HCG is known to be more evolved (e.g., HCG 22) because of low SFRs, large amounts of diffuse \x\ gas, group \hone\ content, or high fraction of ellipticals. 
A complementary way of comparing SFRs and stellar masses for sCGs and HCGs is to view them in the SFR vs. \mstar\ plane, shown in \fr{fig-sfms}. In this parameter space recent work has established a roughly linear correlation between log(SFR) and log(\mstar) known as the ``star forming main sequence" \citep[SFMS;][and references therein]{speagle2014}, and shown within $\pm1\sigma$ by the straight lines in the figure. The correlation holds in both the local and the high-redshift universe (with an offset in normalization) and has also been extracted from cosmological simulations \citep[e.g.,][]{sparre2015}. HCG galaxies as a whole appear to follow the correlation, although they show considerable scatter and nonactively star forming HCGs fall systematically below it. It is striking that our identified sCGs as a whole do not follow the correlation. In particular, there are no sCG galaxies with \mstar~$\gtrsim10^{11}$~\msun, while there is a tail of low-SFR, low-\mstar\ galaxies below the correlation. 

\vspace{.1cm}
On the other hand, separations between pairs of member galaxies are systematically smaller for HCGs. This is so even when we correct for the fact that we can only measure projected separations for HCGs.
We carried out Monte Carlo simulations randomly positioning galaxies in 3D space and then measured both 2D and 3D separations. We deduced that the median 3D distance is a factor of $2\pi/5$ greater than the median projected distance. In the right panel of \fr{fig-hists} we plot sCG pairwise separations against both projected and corrected HCG pairwise separations, showing that sCGs have systematically larger separations by up to more than an order of magnitude in the most extreme cases.

\vspace{0.2cm}
\subsection{$M_{\ast}$ and SFR Comparisons for Individual sCG--HCG Systems}
\vspace{0.2cm}
We now briefly compare corresponding properties of specific sCG and HCG systems, chosen because they are overall closest in \mstar. We compare corresponding \mstar-ranked galaxies. Note that although for sCGs this ranking is always in alphabetical order, with A the most massive galaxy, for HCGs this is not always so for historical reasons: \citet{hickson1982} labeled CG galaxies by letter in order of decreasing $B$-band brightness. Specifically, HCG 22C is more massive than HCG 22B, HCG 31G is more massive than HCG 31B, and HCG 16C is more massive than HCG 16B.
\vspace{0.2cm}
\subsubsection{sCG~115--HCG~22}
The stellar masses of corresponding \mstar-ranked galaxies show very close agreement, within factors of up to $\sim3$ for this pair of triplets. SFRs for HCG 22 galaxies are larger than for sCG 115 ones at $z=0$ by more than an order of magnitude, but comparable to the values $2-3$ Gyr ago.
\vspace{0.2cm}
\subsubsection{sCG~1119--HCG~61}
The stellar masses for HCG 61 galaxies are all about half an order of magnitude higher than those of the corresponding galaxies of sCG 1119. Compared to corresponding sCG galaxies, the SFR is more than two orders of magnitude higher for HCG A and B, but less than an order of magnitude lower for HCG C.
\vspace{0.2cm}
\subsubsection{sCG~1056--HCG~31}
Although HCG 31 nominally consists of four galaxies, galaxies A and C are in an advanced merging stage forming essentially one morphologically peculiar system. ``Galaxy'' E is likely a tidal feature that overlaps closely with A and C, and so all of the A--C--E complex is often treated as a single entity for measuring quantities such as \mstar\ or SFR. In a somewhat similar fashion, sCG 1056 consists of only three galaxies, but, as can be seen from its merger tree (\fr{fig-sCG1056_3D}), its galaxy A consisted of three galaxies in the past. This three-galaxy system first appears at $z\sim0.62 \ (\sim5.7$ Gyr ago). Two subsequent mergers transform it first into a two-galaxy system at $z\sim0.32 \ (\sim3.5$ Gyr ago) and finally into a single galaxy at $z\sim0.06 \ (\sim0.83$ Gyr ago).
\vspace{0.1cm}
\subsubsection{sCG~1598--HCGs~16 and HCG~48}
These are all four-galaxy systems. HCG 16A agrees in \mstar\ with sCG~1598A within a factor of $\sim$2; however all other HCG galaxy \mstar\ values are higher than for their corresponding sCG galaxies by an order of magnitude or more. The situation for SFRs is similar. In contrast, HCG 48 shows the best agreement (within a factor of less than $\sim2$) in stellar masses between the single four-member sCG and any known four-member HCG. In addition, two of the HCG~48 SFR values are also within factors of a few from their corresponding sCG ones; the exceptions are HCG 48A and HCG 48C.
\vspace{0.1cm}
\subsubsection{sCG~377--HCG~37}
Although \mstar\ values are overall higher for HCG galaxies, this pair represents the closest agreement (factors of a few to less than an order of magnitude in \mstar) between the single sCG with five galaxies and a five-member HCG. Apart from galaxy C, HCG SFRs at $z=0$ are more than an order of magnitude higher than those of sCG galaxies. Instead, they are comparable to sCG SFRs $\sim 5$ Gyr ago.

\section{Discussion}\label{sec-disc}
The \mms\ is a chunk of the full \ms\ with 512 times smaller volume. As a result, the identified sCGs, their cosmological evolution, and their comparisons with HCGs represent interesting but essentially random case studies of CG-like systems identified at $z=0$ in a small part of the galaxy catalog produced by applying the semianalytical model of \citet{guo2011} to the full \ms.
We are unable, and do not attempt, to draw any statistically robust conclusions regarding the frequency of occurrence of CGs, whether similar or not to observed systems; this is well beyond the scope of this analysis. That said, the variety of cosmological evolutionary paths identified suggests that there are several ways of leading to observed systems collectively known as CGs. In spite of differences from observed HCGs, it is clear from their merger tree evolution that in sCGs galaxies interact and influence each other, and eventually become bound. The essence of what we understand as a CG is thus an inherent characteristic of these systems.

Compared to HCGs, sCGs are selected differently, the essential initial criterion being proximity within a 3D volume (Section~\ref{sec-clues}) with isolation applied subsequently.  In contrast, the initial selection of HCGs used surface brightness and isolation criteria on galaxy groupings projected on the sky. In addition, the simulation volume is limited, and even more so for the particular catalog used. Even so, we can attempt to formally estimate how many ``HCG-like" systems one might expect to detect as follows. The median redshift of the 92 HCGs with ``accordant" galaxy velocities (spectroscopically verified to be within $\pm1000$ \kmps\ of the group median) is $z=0.03$. %
Using this redshift value, over 67\%\ of the sky \citet{hickson1992} detected 47 CGs with accordant velocities. 
The true number of CGs over 67\% of the sky to this redshift is then $N_{\rm HCG} = 47/f$,
 where $f$ is a completeness factor equal to unity in the ideal case of a catalog with no incompleteness.
Further, the luminosity distance and comoving volume to this redshift are $D_L = 92.1 \, h^{-1}$~Mpc and $V=(\Delta \Omega/3) (D_L/(1+z))^3$,
respectively, where
$\Delta\Omega = 8\pi/3$ is the solid angle subtended by 67\%\ of the sky. The expected number density of HCGs to $z=0.03$ is then 
$n_{\rm HCG}=2.3\times10^{-5} \, h^3 \, f^{-1}{\rm Mpc}^{-3}$.
The volume of a snapshot of the milli-Millennium Simulation is $V_{\rm snap} = (62.5~h^{-1}{\rm Mpc})^3 = 2.44\times 10^5 \, h^{-3} $~Mpc$^3$, and there is one snapshot at $z=0$.  Within this volume, one would expect to find $N_{\rm sim} \sim 6 f^{-1}$ HCGs at $z=0$.  If the sCG volume density were the same as the true volume density of HCGs, our detection of 10 systems might suggest a completeness factor of $\sim 55\%$ for HCGs.  However, the differences in selection criteria between 3D space sCGs and HCGs projected on the sky are significant. In addition, the completeness factor estimated from simulations can be as low as 8\%\ \citep{diaz-gimenez2010}, %
depending on the method and semianalytic model used
%REF
(see \citealt{knebe2015,knebe2018} for a thorough discussion of the effects of different semianalytic models; see also \citealt{snaith2011}). 

Overall, the property that shows the best agreement between HCGs and sCGs is stellar mass. This is consistent with the fact that the \citet{guo2011} semianalytical model uses the observed stellar mass function of galaxies as a primary constraint on several\textit{} parameters related to star formation and feedback (their Table 1). However, when we compare the simulated versus observed CGs in greater detail, we see that, unlike known HCGs, we do not detect sCG galaxies with \mstar~$\gtrsim 10^{11}$~\msun. In contrast, we detect several low to extremely low SFR and \mstar\ sCG galaxies (Figures~\ref{fig-hists} and \ref{fig-sfms}). 
A lack of high-\mstar\ galaxies is perhaps not surprising, given the small simulation volume, as such galaxies are rarer. Note that detection of these galaxies is a direct result of semianalytical rather than full hydrodynamic modeling, so buildup of mass related to intragroup gas, which is known to play a significant role in CGs, may not be properly taken into account. On the other hand, detection of more low-\mstar\ systems is consistent with the fact that the \citet{guo2011} model somewhat overpredicts the abundance of galaxies in the \mstar\ range of our detected sCGs (their Fig.~7), and even more so for low SFRs. The systematically lower SFRs for \mmsg\ groups (sCGs) compared to observations are consistent with a more general trend for lower SFRs in \citet{guo2011} (their Fig. 22 and associated discussion). In addition, our selection is done in 3D space, which is not fully equivalent to the selection of HCGs in projection. Furthermore, it is observationally much harder to detect extremely low SFR systems (which would be fainter in the optical images used for selection by \citealt{hickson1982} than more actively star-forming systems); relatively few studies targeting dwarf galaxy populations have been carried out for individual CGs, yielding few, mostly spectroscopically unconfirmed results (e.g. \citealt{ribeiro1994,krusch2006,konstantopoulos2012,konstantopoulos2013,ordenes-briceno2016,shi2017}, but see \citealt{zabludoff1998,carrasco2006,darocha2011}). This may explain the low-SFR tail obtained for sCGs, combined with the dearth of massive galaxies. No strong conclusion can be made in any case, as we do not have a complete simulation sample in any statistical sense. That said, qualitatively we do observe similar behavior in \mstar\ vs. SFR space, where for the same sCG one or more galaxies may be within 1$\sigma$ of the observed correlation, while one or more may lie below.

We note that our \clues\ selection is based on separation and isolation criteria in 3D. Thus, in a sense sCGs are guaranteed to represent physically compact and isolated systems, perhaps more so than observed HCGs.  With no constraints on relative velocity, however, they are not guaranteed to be bound.  Nonetheless, the fact that velocity vectors stay largely aligned and the systems are isolated suggests that they are expected to be long-lived, unlike, for example, Stephan's Quintet (HCG 92).  HCG~92 has a galaxy, NGC~7318B, that is likely only transiently interacting with the rest of the HCG~92 member galaxies.  Taken at face value, the long lifetimes of the $z=0$ sCGs (see Columns (8) and (10) in Table~1) support the idea that at least some CGs are not transient concentrations.  Such examples bring to mind the local example of HCG~7, a quartet of galaxies that appear to be relatively undisturbed morphologically, but whose properties such as H~{\sc i} mass and distribution and population of globular clusters indicate an extended period of group evolution without evidence for a major merger \citep{konstantopoulos2010}.

Finally, although observed HCGs appear to be systematically more compact than sCGs, the discrepancy in pairwise galaxy separations will at least partially be affected by projection effects. For instance, sCG triplets evolve along roughly parallel paths in 3D. Depending on viewing angle, these can be projected on the sky to appear almost as compact as HCGs.

\section{Summary and Conclusions}\label{sec-summ}
We used the \clues\ algorithm to identify and characterize 10 CGs of galaxies in a portion of the semianalytical galaxy catalog of \citet{guo2011}, corresponding to the milli-Millennium cosmological simulation (sCGs in the \mmsg).  We further compared key properties of these to those of observed HCGs. This work provides an independent cross-check of the Guo2010a semianalytical model by probing a regime that the model was not tuned to. 
Our main conclusions are as follows:
\begin{enumerate}
\itemsep-4pt 
\item With sCG stellar masses in the range $10^{10} \lesssim M_{\ast}/M_{\odot} \lesssim 10^{11}$, \mstar\ is the property for which the greatest number (six) of HCGs have at least one galaxy that agrees within factors of a few with sCG galaxies at $z=0$. However, most (76\%) HCG galaxies have higher \mstar\ than sCG ones.
\item Once sCG galaxies come within 1 (0.5) Mpc of their most massive galaxy, they remain within that distance until $z=0$.  The typical redshifts for these ``birthdays'' indicate that $z=0$ sCGs are long-lived systems, with ages in the range of 5.7--11.4~Gyr (1.1--8.9~Gyr for the $\Delta D_{\rm A}=0.5$~Mpc criterion).
\item The pairwise separations of sCG galaxies show a variety of cosmological evolution. Although some exhibit a roughly oscillatory behavior, we identify one system where member galaxies consistently come closer to each other, a clear candidate for merging into a single galaxy or becoming a ``fossil" group.
\item We define the angular pairwise 3D velocity vector separation, \dtheta, to capture the qualitative distinction between galaxies that approach on trajectories that can be characterized as ``collision courses'' (e.g., sCG 1598; Fig. \ref{fig-sCG1598_3D}) vs. galaxies on more parallel tracks (e.g., sCG 1056; Fig. \ref{fig-sCG1056_3D}).  We speculate that the latter are expected to be more long-lived, with fewer instances of major mergers.
\item Triplet galaxies show the smallest changes in pairwise separations after close encounters and appear to travel across more parallel paths over cosmic times compared to nontriplets, thus making it more likely that they are not transient systems.
\item Only some sCG galaxies follow the SFR--\mstar\ correlation, with no galaxies with \mstar~$\gtrsim 10^{11}$~\msun\ and a tail of low SFR and \mstar.
\item At $z=0$, HCG galaxies have systematically higher SFRs than sCG members by up to more than an order of magnitude.
\item With a few exceptions, pairwise galaxy separations are systematically smaller for HCGs; it is unclear how important projection effects are in this respect.  
\end{enumerate}

Given the small volume of the simulation and the handful of identified sCGs, no results of a statistical nature can be drawn. Our results are case studies of possible pathways to CGs at $z=0$. The probability of these particular pathways in the real universe remains unknown, while other pathways are likely possible. These results should be considered as a launching point for further, more detailed work, using larger and more recent simulations in tandem with observational catalogs.
\acknowledgments
We thank the anonymous referee for their constructive comments and appreciation of this work.

Authorship statement: S.A., D.R.M., and S.P. contributed substantially and equally to the technical analysis in this paper; their order in the author list is therefore alphabetical and does not indicate their relative contribution of effort.

P.T. acknowledges support from NASA grant NNX15AK25G (solicitation NNH14ZDA001N-ADAP). S.C.G., S.A., and D.R.M. acknowledge support from the Natural Sciences and Engineering Research Council of Canada. K.E.J. is grateful to the David and Lucile Packard Foundation for their generous support. 

{\it Software}: astropy \citep{astropy2013}, R \citep{rteam2012}.

\begin{center}
  {\bf ORCID iDs}
\end{center}

\noindent P.~Tzanavaris\\
\noindent K.~E.~Johnson

\bibliographystyle{likeapj}
\bibliography{masterbib,xlc,extra}   

\end{document}